\documentclass[11pt,twoside]{article}
\input{epsf.sty}
\usepackage{mathrsfs}
\usepackage{amsmath}
\usepackage{amssymb}
\usepackage{graphicx}
\usepackage{subfigure}
\usepackage{cite}
\newtheorem{definition}{Definition}
\newtheorem{theorem}{Theorem}
\newtheorem{proposition}{Proposition}

\numberwithin{equation}{section}
\newcommand*{\QEDB}{\hfill\ensuremath{\square}}

\title{The Ablowitz-Ladik system on a finite set of integers}
\author{Baoqiang Xia
\\
School of Mathematics and Statistics, Jiangsu Normal
University,\\
 Xuzhou, Jiangsu 221116, P. R. China,
 \\
 E-mail address:
xiabaoqiang@126.com}

\date{}
\begin{document}
\maketitle
\begin{abstract}
We show how to solve initial-boundary value problems for integrable nonlinear differential-difference equations on a finite set of integers.
The method we employ is the discrete analogue of the unified transform (Fokas method).
The implementation of this method to the Ablowitz-Ladik system yields the solution in
terms of the unique solution of a matrix Riemann-Hilbert problem,
which has a jump matrix with explicit $(n,t)$-dependence involving certain functions referred to as spectral functions.
Some of these functions are defined in terms of the initial value, while the remaining spectral functions are defined in terms of two sets
of boundary values. These spectral functions are not independent but satisfy an algebraic relation called global relation.
We analyze the global relation to characterize the unknown boundary values in terms of the given
initial and boundary values. We also discuss the linearizable boundary conditions.


\end{abstract}
\newpage

\section{ Introduction}

The so-called unified transform \cite{F1,F6},
which is also referred to as the Fokas method,
provides a general way for solving initial-boundary value problems (IBVPs) for integrable partial differential equations (PDEs).
By performing the simultaneous spectral analysis of the Lax pair associated with a PDE,
the method expresses the solution in terms of the solution of a matrix Riemann-Hilbert (RH) problem formulated in the complex plane.
The Fokas method has been extensively used in the literature to analyse the IBVPs for integrable PDEs on the half line, on the interval and also on simple graph
structures; see for example \cite{F3,MFS,F4,F5,F6,F7,MFS1,LF1,L1,L2,XF,XF2,GLZ,AK,C}. In addition to IBVPs for integrable PDEs, another natural and important issue is the study of IBVPs for integrable differential-difference equations (DDEs).
Recently, Biondini and collaborators \cite{BH1,BH2,BH3} have initiated the study of implementing the Fokas method to study IBVPs for integrable DDEs on natural numbers.
The method of Biondini and collaborators differs from the analogous analysis in PDEs case in two important ways:
first, the determinants of the eigenfunctions of the $n$-part of the Lax pair depend on the potential and the independent variables;
second, the $t$-part of the Lax pair is not traceless. These difficulties were overcame in our very recent paper \cite{XiaFokas}.

In the present paper, we show how to solve IBVPs for integrable nonlinear DDEs on a finite set of integers.
The illustrative example we consider is the integrable discrete nonlinear Schr\"{o}dinger (DNLS) equation \cite{AL1,AL2,AL3}
on a finite set of integers
\begin{eqnarray}
i\frac{dq_n}{dt}+q_{n+1}-2q_{n}+q_{n-1}-\nu |q_n|^2\left(q_{n+1}+q_{n-1}\right)=0, ~\nu=\pm 1,~0\leq n\leq N-1, ~0<t<T,
\label{al1}
\end{eqnarray}
with the given initial and boundary values
\begin{eqnarray}
\begin{split}
&q(n,0)=q_{0}(n), \quad 0\leq n\leq N-1,
\\
&q(-1,t)=g_{-1}(t), \quad q(N,t)=g_N(t), \quad 0<t<T.
\end{split}
\label{ibvp}
\end{eqnarray}
Here $q_{0}(n)$, $g_{-1}(t)$, $g_N(t)$ are compatible at $n = t = 0$
and at $n = N$, $t = 0$, i.e. $q_{0}(-1)= g_{-1}(0)$, $q_{0}(N)= g_N(0)$.
The DNLS equation also referred to as Ablowitz-Ladik lattice system is an important integrable discrete model with a number of physical and mathematical contexts \cite{AL1,AL2,AL3}.
Our analysis 
is based on the extension of the results of \cite{XiaFokas} from the non-negative integers to the finite set of integers.
In analogy with the integrable PDEs on the interval, the analysis involves the following three steps.

{\it Step 1. A RH formulation under the assumption of existence.} We assume that there exists a solution $q(n,t)$.
By performing the simultaneous spectral analysis of the associated Lax pair of the DNLS equation, we express $q(n,t)$ in terms of the solution of a $2 \times 2$ matrix RH problem
defined in the complex $z$-plane. This RH problem has explicit $(n,t)$-dependence in the form of
$z^{2n}e^{i(z-z^{-1})^2t}$, and it is uniquely defined in terms of the  so-called spectral functions
$\left\{a(z),b(z)\right\}$, $\left\{A(z),B(z)\right\}$ and $\left\{A_N(z),B_N(z)\right\}$.
The spectral functions $\left\{a(z),b(z)\right\}$ depend on the initial value: $q_{0}(n)$;
whereas the spectral functions $\left\{A(z),B(z)\right\}$ and $\left\{A_N(z),B_N(z)\right\}$ depend on two sets of boundary values: $\{g_{-1}(t),g_0(t)\}$ and $\{g_{N-1}(t), g_N(t)\}$, where $g_{0}(t)$ and $g_{N-1}(t)$ denote the unknown boundary values $q(0,t)$ and $q(N-1,t)$.
 We show that the spectral functions are not independent but they satisfy an algebraic relation called global relation.

{\it Step 2. Existence under the assumption that the spectral functions satisfy the global relation.}
We define the spectral functions $\{a(z),b(z)\}$ in terms of the
initial datum $q_0(n)$,
define the spectral functions $\{A(z),B(z)\}$ in terms of the boundary values $g_{-1}(t)$ and $g_{0}(t)$, and
define the spectral functions $\{A_N(z),B_N(z)\}$ in terms of the boundary values $g_{N-1}(t)$ and $g_{N}(t)$.
We assume that the boundary values are such that the spectral functions satisfy
the global relation. We also define $q(n,t)$ in terms of the solutions of the RH problems
formulated in step 1. We then prove that the formula for $q(n,t)$ solves the DNLS equation,
and furthermore satisfies the given initial and boundary conditions, i.e. $q(n,0)=q_0(n)$,
$q(-1,t)=g_{-1}(t)$ and $q(N,t)=g_{N}(t)$.

{\it Step 3. Elimination of the unknown boundary values.}
Given $q_0(n)$, $g_{-1}(t)$, $g_{N}(t)$, by employing the global relation, we characterize the unknown boundary value $g_0(t)$ and $g_{N-1}(t)$
through the solution of a system of nonlinear Volterra integral equations.

The paper is organized as follows: in sections \ref{section2}-\ref{section4} we implement steps 1-3.
In section 5, we present a particular class of boundary conditions, called linearizable boundary conditions.
Our results are discussed further in section 6.


\section{A Riemann-Hilbert formulation under the assumption of existence}\label{section2}

Assume a solution $q(n,t)$ exists, we will express $q(n,t)$ in terms of the solution of a $2\times 2$ matrix RH problem
defined in the complex $z$-plane.
To achieve this, we perform the simultaneous spectral analysis of the associated Lax pair
by considering appropriate solutions of the $n$-part of the Lax pair evaluated at $t=0$ and of the $t$-part
of the Lax pair evaluated at $n=0$ and at $n=N$.

\subsection{A Lax pair}

We consider a pair of linear spectral problem
\begin{subequations}
\begin{align}
&\Phi(n+1,t,z)-\frac{1}{f(n,t)}Z\Phi(n,t,z)=
\frac{1}{f(n,t)}Q(n,t)\Phi(n,t,z),
\label{LPS}\\
&\Phi_t(n,t,z)-i\omega(z)\sigma_3\Phi(n,t,z)=
H(n,t,z)\Phi(n,t,z),
\label{LPT}
\end{align}
\label{LP}
\end{subequations}
where $\Phi(n,t,z)$ is a $2\times 2$ matrix-valued function, and
\begin{eqnarray}
\begin{split}
&f(n,t)=\sqrt{1-\nu |q(n,t)|^2},\quad \omega(z)=\frac{1}{2}(z-z^{-1})^2,
\\
&Z=\left( \begin{array}{cc} z & 0 \\
 0 &  z^{-1}\\ \end{array} \right),
 \quad
\sigma_3=\left( \begin{array}{cc} 1 & 0 \\
 0 &  -1 \\ \end{array} \right),
 \quad
Q(n,t)=\left( \begin{array}{cc} 0 & q(n,t) \\
 \nu q^\ast(n,t) &  0 \\ \end{array} \right),
 \\
&H(n,t,z)=i\left( \begin{array}{cc} -\nu \text{Re} (q(n,t)q^\ast(n-1,t))&  zq(n,t)- z^{-1}q(n-1,t) \\  \nu(zq^\ast(n-1,t)- z^{-1}q^\ast(n,t)) & \nu \text{Re} (q(n,t)q^\ast(n-1,t)) \\ \end{array} \right).
\end{split}
\label{QH0}
\end{eqnarray}
Hereafter we use the symbols $\text{Re}(\cdot)$ and $\text{Im}(\cdot)$ denote the real and imaginary parts of a complex-valued function.
It can be checked directly that the compatibility condition of (\ref{LPS}) and (\ref{LPT}) yields nothing but the DNLS equation (\ref{al1}).
Thus (\ref{LP}) provides a Lax pair formulation for the DNLS equation (\ref{al1}).
We note that the $n$-part of Lax pair (\ref{LPS}) has already appeared in \cite{Miller,Geng1}.

The Lax pair (\ref{LP}) in comparison with the one in \cite{BH1,AL1} possesses the following two properties:
first, the matrix $U(n,t)=\frac{1}{f(n,t)}(Z+Q(n,t))$ appearing in the $n$-part of the Lax pair is unimodular;
second, the matrix $V(n,t)=i\omega(z)\sigma_3+H(n,t,z)$ appearing in the $t$-part of the Lax pair is traceless.
It will become clear in section \ref{sec2.3} that these two properties are very convenient for performing spectral analysis
and for formulating the relevant RH problem.

We define the modified eigenfunction $\mu(n,t,z)$ by
\begin{eqnarray}
\Phi(n,t,z)=\mu(n,t,z)Z^ne^{iw(z)t\sigma_3}.
\label{mud}
\end{eqnarray}
Then the Lax pair (\ref{LP}) becomes
\begin{subequations}
\begin{align}
&\mu(n+1,t,z)-\frac{1}{f(n,t)}\hat{Z}\mu(n,t,z)=
\frac{1}{f(n,t)}Q(n,t)\mu(n,t,z)Z^{-1},
\label{LPS3}\\
&\mu_t(n,t,z)-i\omega(z)[\sigma_3,\mu(n,t,z)]=
H(n,t,z)\mu(n,t,z),
\label{LPT3}
\end{align}
\label{LP3}
\end{subequations}
where $\hat{Z}$ acts on a $2 \times 2$ matrix $A$ as follows:
\begin{eqnarray}
\hat{Z}A=ZAZ^{-1}.
\label{zhatd1}
\end{eqnarray}
Furthermore, we let
\begin{eqnarray}
\Psi(n,t,z)=\hat{Z}^{-n}e^{-iw(z)t\hat{\sigma_3}}\mu(n,t,z),
\label{mud}
\end{eqnarray}
where $e^{\hat{\sigma_3}}$ acts on a $2 \times 2$ matrix $A$ as follows:
\begin{eqnarray}
e^{\hat{\sigma_3}}A=e^{\sigma_3}Ae^{-\sigma_3}.
\label{zhatd2}
\end{eqnarray}
Then the Lax pair (\ref{LP3}) becomes
\begin{subequations}
\begin{align}
&\Psi(n+1,t,z)-\frac{1}{f(n,t)}\Psi(n,t,z)=
\frac{1}{f(n,t)}Z^{-1}\hat{Z}^{-n}e^{-iw(z)t\hat{\sigma_3}}\left(Q(n,t)\right)\Psi(n,t,z),
\label{LPS4}\\
&\Psi_t(n,t,z)=\hat{Z}^{-n}e^{-iw(z)t\hat{\sigma_3}}\left(H(n,t,z)\right)\Psi(n,t,z).
\label{LPT4}
\end{align}
\label{LP4}
\end{subequations}

It is convenient to introduce the following notations
\begin{eqnarray}
\begin{split}
&C(n,t)=\prod^{N-1}_{m=n}f(m,t),~ 0\leq n\leq N-1, \quad C(N,t)=1,
\\
&E(-1,t)=\exp\left(-\nu \int_0^t \text{Im}\left(g_{0}(t') g^*_{-1}(t')\right)dt'\right),
\\
&E(N-1,t)=\exp\left(-\nu \int_0^t \text{Im}\left(g_{N-1}(t') g^*_{N}(t')\right)dt'\right),
\\
&\hat{E}(-1,t)=\exp\left(\nu \int_t^T \text{Im}\left(g_{0}(t') g^*_{-1}(t')\right)dt'\right),
\\
&\hat{E}(N-1,t)=\exp\left(\nu \int_t^T \text{Im}\left(g_{N-1}(t') g^*_{N}(t')\right)dt'\right),
\end{split}
\label{nota1}
\end{eqnarray}
By employing equation (\ref{al1}) we can compute the $t$-derivative of $C(n,t)$:
\begin{eqnarray}
C_t(n,t)=-\nu \text{Im}\left(q(n,t) q^\ast(n-1,t)+ g_{N-1}(t)g_{N}^\ast(t)\right)C(n,t), ~ 0\leq n\leq N-1.
\label{ct}
\end{eqnarray}
Hence,
\begin{eqnarray}
\frac{C(n,t)}{C(n,0)}=\exp\left(-\nu \int_0^t \text{Im}\left(q(n,t') q^\ast(n-1,t')+ g_{N-1}(t')g^*_{N}(t')\right)dt'\right), 0\leq n\leq N-1.
\label{cts}
\end{eqnarray}
In particular, for $n=0$ we find
\begin{eqnarray}
\frac{C(0,t)}{C(0,0)}=E(-1,t)E(N-1,t).
\label{ct1}
\end{eqnarray}

\subsection{The eigenfunctions}\label{sec2.2}

\begin{figure}
\centering
\includegraphics[width=4.0in,height=3.0in]{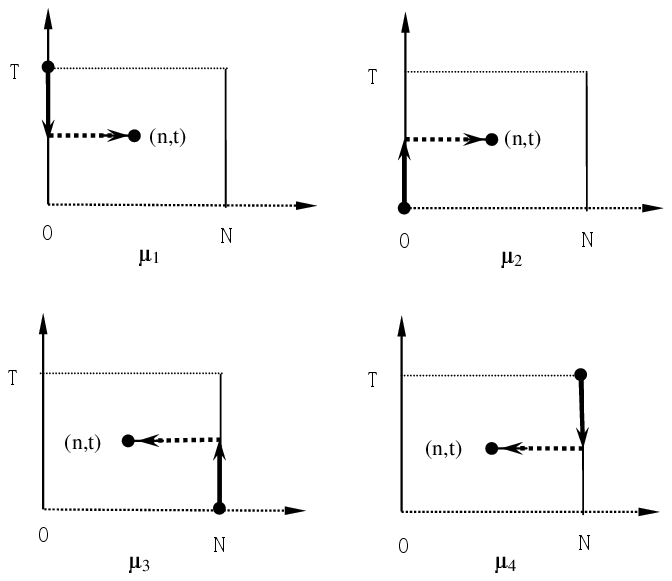}
\caption{\small{ The contours used for the definition of $\left\{\mu_j\right\}_{1}^4$}}
\label{F1}
\end{figure}

As in the continuous case, we define four eigenfunctions
with normalizations at each of the corners of the polygonal domain \cite{F8}.
More precisely, making use of the modified eigenfunction $\Psi(n,t,z)$ in (\ref{LP4}),
we define four eigenfunctions $\{\mu_j(n,t,z)\}_{1}^4$ that approach the identity matrix
respectively at $(n,t)=(0,T)$, at $(n,t)=(0,0)$, at $(n,t)=(N,0)$, and at $(n,t)=(N,T)$ (see figure 1) as follows:
\begin{eqnarray}
\begin{split}
\mu_{1}(n,t,z)=&\frac{C(n,t)}{C(0,t)}\left(I-\hat{Z}^{n}\int_{t}^{T}e^{iw(z)(t-t')\hat{\sigma}_3}\left(H\mu_{1}(0,t',z)\right)dt'\right)
\\&+C(n,t)Z^{-1}\sum_{m=0}^{n-1} \frac{1}{C(m,t)}\hat{Z}^{n-m}(Q(m,t)\mu_{1}(m,t,z)),
 \\
\mu_{2}(n,t,z)=&\frac{C(n,t)}{C(0,t)}\left(I+\hat{Z}^{n}\int_{0}^{t}e^{iw(z)(t-t')\hat{\sigma}_3}\left(H\mu_{2}(0,t',z)\right)dt'\right)
\\&+C(n,t)Z^{-1}\sum_{m=0}^{n-1} \frac{1}{C(m,t)}\hat{Z}^{n-m}(Q(m,t)\mu_{2}(m,t,z)),
\\
\mu_{3}(n,t,z)=&\frac{1}{C(n,t)}\left(I+\hat{Z}^{-(N-n)}\int_{0}^{t}e^{iw(z)(t-t')\hat{\sigma}_3}\left(H\mu_{3}(N,t',z)\right)dt'\right)
\\&-\frac{1}{C(n,t)}\sum_{m=n+1}^{N}C(m,t)\hat{Z}^{-(m-n-1)}(Q(m-1,t)\mu_{3}(m,t,z))Z,
\\
\mu_{4}(n,t,z)=&\frac{1}{C(n,t)}\left(I-\hat{Z}^{-(N-n)}\int_{t}^{T}e^{iw(z)(t-t')\hat{\sigma}_3}\left(H\mu_{4}(N,t',z)\right)dt'\right)
\\&-\frac{1}{C(n,t)}\sum_{m=n+1}^{N}C(m,t)\hat{Z}^{-(m-n-1)}(Q(m-1,t)\mu_{4}(m,t,z))Z.
\end{split}
\label{muibv}
\end{eqnarray}
Note that all the $\mu_{j}(n,t,z)$ are analytic in the punctured complex $z$-plane $\mathbb{C}/\{0\}$.

We introduce the following domains (see figure 2):
\begin{eqnarray}
\begin{split}
&D_{-in}=\left\{z\Big| z\in\mathbb{C}, |z|<1, \arg z\in(\frac{\pi}{2},\pi)\cup (\frac{3\pi}{2},2\pi)\right\},
\\
&D_{+in}=\left\{z\Big| z\in\mathbb{C}, |z|<1, \arg z\in(0,\frac{\pi}{2})\cup (\pi,\frac{3\pi}{2})\right\},
\\
&D_{-out}=\left\{z\Big| z\in\mathbb{C}, |z|>1, \arg z\in(0,\frac{\pi}{2})\cup (\pi,\frac{3\pi}{2})\right\},
\\
&D_{+out}=\left\{z\Big| z\in\mathbb{C}, |z|>1, \arg z\in(\frac{\pi}{2},\pi)\cup (\frac{3\pi}{2},2\pi)\right\},
\\
&D_{in}=\left\{z\Big| z\in\mathbb{C}, |z|<1 \right\}, ~~D_{out}=\left\{z\Big| z\in\mathbb{C}, |z|>1 \right\},
\\
&D_{-}=\left\{z\Big| z\in\mathbb{C}, \text{Im} (\omega(z))>0\right\}, ~~D_{+}=\left\{z\Big| z\in\mathbb{C}, \text{Im} (\omega(z))<0\right\}.
\end{split}
\label{D1}
\end{eqnarray}
Let $\{\mu^L_j(n,t,z)\}_{1}^4$ and $\{\mu^R_j(n,t,z)\}_{1}^4$ denote the first and second column of $\{\mu_j(n,t,z)\}_{1}^4$ respectively.
These columns are bounded in the following domains of the complex $z$-plane:
\begin{eqnarray}
\begin{split}
&\mu_{1}^L(n,t,z): \bar{D}_{-out},\quad \mu_{1}^R(n,t,z): \bar{D}_{+in},
\\
&\mu_{2}^L(n,t,z):\bar{D}_{+out},\quad \mu_{2}^R(n,t,z): \bar{D}_{-in},
\\
&\mu_{3}^L(n,t,z): \bar{D}_{+in},\quad \mu_{3}^R(n,t,z): \bar{D}_{-out},
\\
&\mu_{4}^L(n,t,z): \bar{D}_{-in},\quad \mu_{4}^R(n,t,z): \bar{D}_{+out},
\end{split}
\label{DE}
\end{eqnarray}
where $\bar{D}$ denotes the closure of a domain $D$.

By using (\ref{muibv}) and employing Neumann series (for details see appendix A in \cite{XiaFokas}),
we can derive the asymptotic behavior of $\mu_j(n,t,z)$ both as $z\rightarrow 0$ and as $z\rightarrow \infty$.
For $1\leq n\leq N-1$, we find
\begin{eqnarray}
\begin{split}
\mu_1(n,t,z)&=\frac{C(n,t)}{C(0,t)}\hat{E}(-1,t)\left(I+Q(n-1,t)Z^{-1}+\left( \begin{array}{cc} O(z^{-2},\text{even}) & O(z^{3},\text{odd})  \\
  O(z^{-3},\text{odd}) & O(z^{2},\text{even}) \\ \end{array} \right)\right), ~ z\rightarrow (\infty,0),
\\
\mu_2(n,t,z)&=\frac{C(n,t)}{C(0,t)}E(-1,t)\left(I+Q(n-1,t)Z^{-1}+\left( \begin{array}{cc} O(z^{-2},\text{even}) & O(z^{3},\text{odd})  \\
  O(z^{-3},\text{odd}) & O(z^{2},\text{even}) \\ \end{array} \right)\right), ~ z\rightarrow (\infty,0),
\\
  \mu_3(n,t,z)&=\frac{E(N-1,t)}{C(n,t)}\left(I-Q(n,t)Z+\left( \begin{array}{cc} O(z^{2},\text{even}) & O(z^{-3},\text{odd})  \\
  O(z^{3},\text{odd}) & O(z^{-2},\text{even}) \\ \end{array} \right)\right), \quad z\rightarrow (0,\infty),
\\
\mu_4(n,t,z)&=\frac{\hat{E}(N-1,t)}{C(n,t)}\left(I-Q(n,t)Z+\left( \begin{array}{cc} O(z^{2},\text{even}) & O(z^{-3},\text{odd})  \\
  O(z^{3},\text{odd}) & O(z^{-2},\text{even}) \\ \end{array} \right)\right), \quad z\rightarrow (0,\infty),
\end{split}
\label{asp1}
\end{eqnarray}
where the functions $C(n,t)$, $E(-1,t)$, $E(N-1,t)$, $\hat{E}(-1,t)$, $\hat{E}(-1,t)$ are defined by (\ref{nota1}).
Here and in what follows the notation $O(z^{2},even)$ ($O(z^{-2},even)$) means that there is a convergent
asymptotic expansion of the remainder in $z$ ($z^{-1}$) and the higher order terms are all even
powers of $z$ ($z^{-1}$),
while the notation $O(z^{3},\text{odd})$ ($O(z^{-3},\text{odd})$) means the higher order terms of the expansion are all odd
powers of $z$ ($z^{-1}$), and the notation $z\rightarrow (\infty,0)$ ($z\rightarrow (0,\infty)$) means $z\rightarrow \infty$ ($z\rightarrow 0$) for the first column of $\mu_j(n,t,z)$
and $z\rightarrow 0$ ($z\rightarrow \infty$) for the second column of $\mu_j(n,t,z)$.
For $n=0$, we find
\begin{eqnarray}
\begin{split}
\mu_1(0,t,z)=&\hat{E}(-1,t)I+\left(\hat{E}(-1,t)Q(-1,t)-e^{iw(z)(t-T)\hat{\sigma}_3}Q(-1,T)\right)Z^{-1}
\\&+\left( \begin{array}{cc} O(z^{-2},\text{even}) & O(z^{3},\text{odd})  \\
  O(z^{-3},\text{odd}) & O(z^{2},\text{even}) \\ \end{array} \right), \quad z\rightarrow (\infty,0),
\\
\mu_2(0,t,z)=&E(-1,t)I+\left(E(-1,t)Q(-1,t)-e^{iw(z)t\hat{\sigma}_3}Q(-1,0)\right)Z^{-1}
\\&+\left( \begin{array}{cc} O(z^{-2},\text{even}) & O(z^{3},\text{odd})  \\
  O(z^{-3},\text{odd}) & O(z^{2},\text{even}) \\ \end{array} \right), \quad z\rightarrow (\infty,0).
\end{split}
\label{asp2}
\end{eqnarray}
For $n=N$, we find
\begin{eqnarray}
\begin{split}
\mu_3(N,t,z)=&E(N-1,t)I-\left(E(N-1,t)Q(N,t)-e^{iw(z)t\hat{\sigma}_3}Q(N,0)\right)Z
\\&+\left( \begin{array}{cc} O(z^{2},\text{even}) & O(z^{-3},\text{odd})  \\
  O(z^{3},\text{odd}) & O(z^{-2},\text{even}) \\ \end{array} \right), \quad z\rightarrow (0,\infty),
\\
\mu_4(N,t,z)=&\hat{E}(N-1,t)I-\left(\hat{E}(N-1,t)Q(N,t)-e^{iw(z)(t-T)\hat{\sigma}_3}Q(N,T)\right)Z
\\&+\left( \begin{array}{cc} O(z^{2},\text{even}) & O(z^{-3},\text{odd})  \\
  O(z^{3},\text{odd}) & O(z^{-2},\text{even}) \\ \end{array} \right), \quad z\rightarrow (0,\infty).
\end{split}
\label{asp3}
\end{eqnarray}

\begin{figure}
\begin{minipage}[t]{0.5\linewidth}
\centering
\includegraphics[width=2.5in]{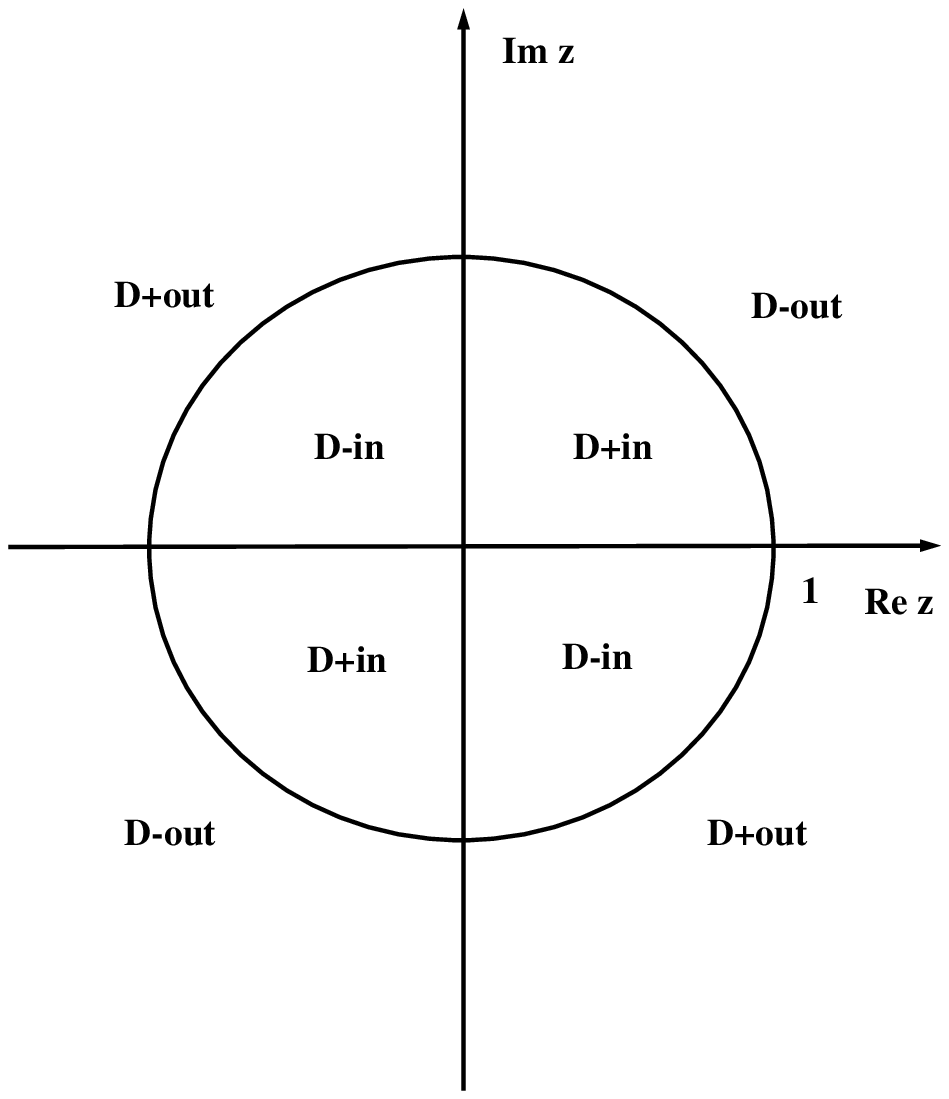}
\caption{\small{ The domains $D_{-in}$, $D_{+in}$, $D_{-out}$ and $D_{+out}$ of the $z$-plane.}}
\label{F1}
\end{minipage}
\hspace{2.0ex}
\begin{minipage}[t]{0.5\linewidth}
\centering
\includegraphics[width=2.5in]{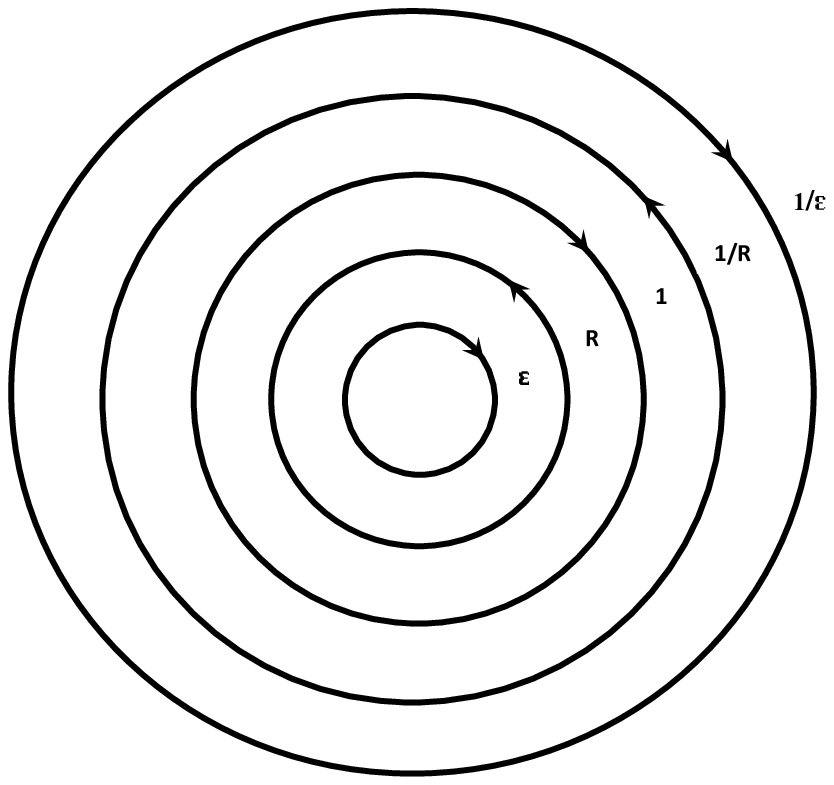}
\caption{\small{ The oriented contours $L^{(n)}$.}}
\label{F2}
\end{minipage}
\end{figure}

\begin{figure}
\centering
\includegraphics[width=4.0in,height=4.0in]{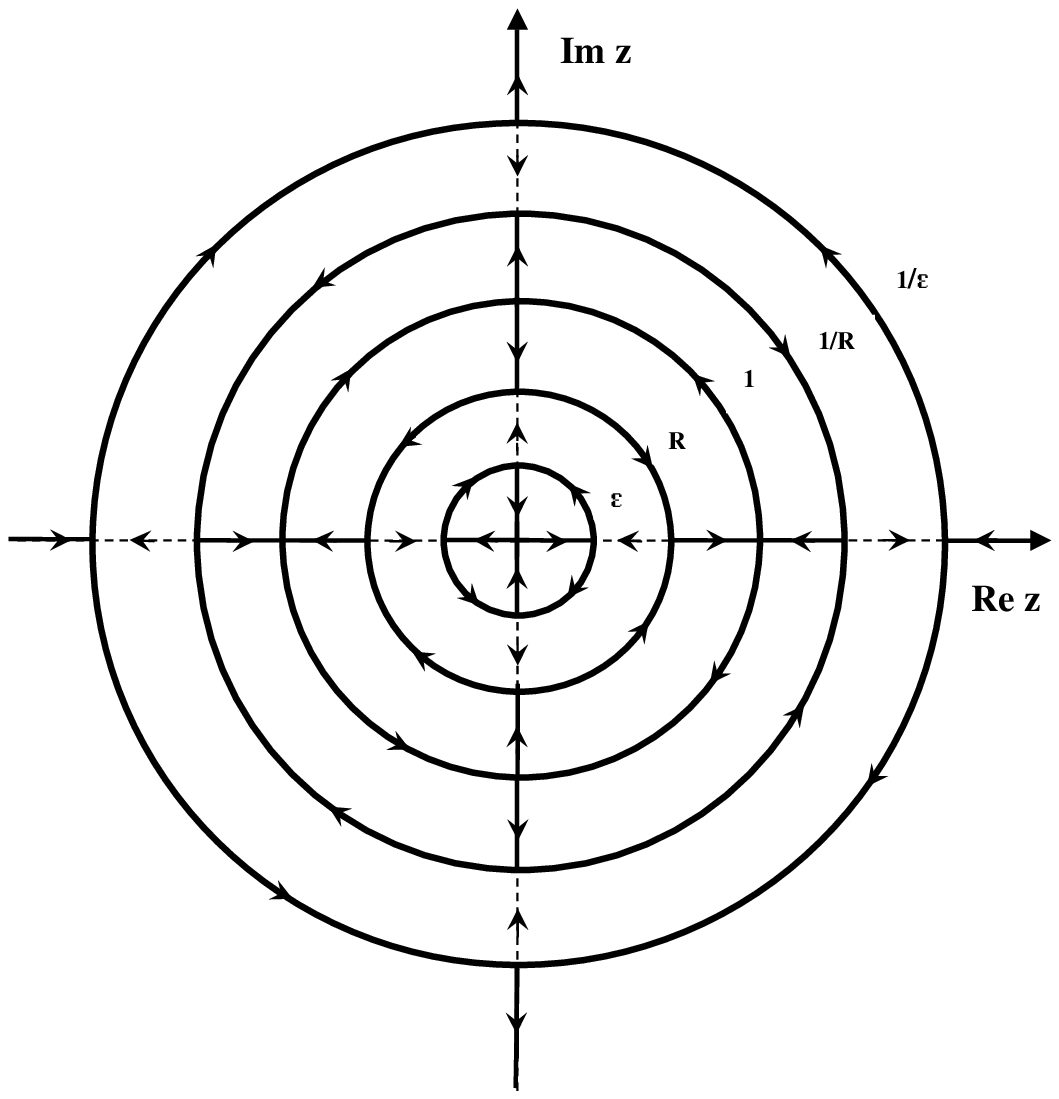}
\caption{\small{ The oriented contours $L$: the jumps across the dashed parts of $L$ are trivial.}}
\label{F3}
\end{figure}

\subsection{The spectral functions}\label{sec2.3}

The $n$-part of Lax pair (\ref{LPS}) implies the equation $\det\Phi(n+1,t,z)=\det\Phi(n,t,z)$.
The matrix $V(n)=i\omega(z)\sigma_3+H(n,t,z)$ appearing in the $t$-part of the Lax pair is traceless.
These two facts imply the important identities:
\begin{eqnarray}
\det\Phi_j(n,t,z)=1, \quad j=1, 2, 3, 4.
\label{detp1}
\end{eqnarray}

Since the matrices $\{\Phi_j(n,t,z)\}_1^4$ are fundamental solutions of the same Lax pair,
they are related by the equations
\begin{subequations}
\begin{align}
\Phi_3(n,t,z)&=\Phi_2(n,t,z)s(z), ~~z\in\mathbb{C}/\{0\},\label{s3a}
\\
\Phi_1(n,t,z)&=\Phi_2(n,t,z)S(z), ~~z\in\mathbb{C}/\{0\},\label{s3b}
\\
\Phi_4(n,t,z)&=\Phi_3(n,t,z)\hat{Z}^{-N}S_N(z),~~z\in\mathbb{C}/\{0\}. \label{s3c}
\end{align}
\label{s3}
\end{subequations}
Then the matrices $\{\mu_j(n,t,z)\}_1^4$ satisfy the equations
\begin{subequations}
\begin{align}
\mu_3(n,t,z)&=\mu_2(n,t,z)\hat{Z}^ne^{i\omega(z)t\hat{\sigma_3}}s(z), ~~z\in\mathbb{C}/\{0\},\label{s4}
\\
 \mu_1(n,t,z)&=\mu_2(n,t,z)\hat{Z}^ne^{i\omega(z)t\hat{\sigma_3}}S(z),~~z\in\mathbb{C}/\{0\},\label{s5}
 \\
\mu_4(n,t,z)&=\mu_3(n,t,z)\hat{Z}^{n-N}e^{i\omega(z)t\hat{\sigma_3}}S_N(z),~~z\in\mathbb{C}/\{0\}.
\label{s6}
\end{align}
\label{s456}
\end{subequations}
Equations (\ref{detp1}) and (\ref{s3}) yield
\begin{eqnarray}
\det s(z)=\det S(z)=\det S_N(z)=1,~~z\in\mathbb{C}/\{0\}.
\label{det2}
\end{eqnarray}
Equations (\ref{s4}) and (\ref{s6}) imply
\begin{eqnarray}
\mu_4(n,t,z)=\mu_2(n,t,z)\hat{Z}^ne^{i\omega(z)t\hat{\sigma_3}}\left(s(z)\hat{Z}^{-N}S_N(z)\right),~~z\in\mathbb{C}/\{0\}.
\label{s7}
\end{eqnarray}

Evaluating equations (\ref{s4}) at $n=t=0$, we find
\begin{eqnarray}
\begin{split}
s(z)&=\mu_3(0,0,z).
\end{split}
\label{sSr11}
\end{eqnarray}
Evaluating equations (\ref{s5}) at $n=t=0$, we find
\begin{eqnarray}
\begin{split}
S(z)&=\mu_1(0,0,z).
\end{split}
\label{sSr12}
\end{eqnarray}
Evaluating equation (\ref{s6}) at $n=N$, $t=0$ we find
\begin{eqnarray}
\begin{split}
S_N(z)&=\mu_4(N,0,z).
\end{split}
\label{sSr13}
\end{eqnarray}
Evaluating equation (\ref{s5}) at $n=0$, $t=T$, we find
\begin{eqnarray}
\begin{split}
S(z)=\left(e^{-i\omega(z)T\hat{\sigma_3}}\mu_2(0,T,z)\right)^{-1}.
\end{split}
\label{sSr2}
\end{eqnarray}
Evaluating equation (\ref{s6}) at $n=N$, $t=T$, we find
\begin{eqnarray}
\begin{split}
S_N(z)&=\left(e^{-i\omega(z)T\hat{\sigma_3}}\mu_3(N,T,z)\right)^{-1}.
\end{split}
\label{sSr3}
\end{eqnarray}
Equations (\ref{muibv}) and (\ref{sSr11}), (\ref{sSr2}), (\ref{sSr3}) imply
\begin{eqnarray}
\begin{split}
s(z)&=C(0,0)\left(I-Z^{-1}\sum_{m=0}^{N-1} \frac{1}{C(m,0)}\hat{Z}^{-m}\left(Q(m,0)\mu_{3}(m,0,z)\right)\right),
\\
S^{-1}(z)&=I+\int_{0}^{T}e^{-iw(z)t\hat{\sigma}_3}\left(H\mu_{2}(0,t,z)\right)dt,
\\
S_N^{-1}(z)&=I+\int_{0}^{T}e^{-iw(z)t\hat{\sigma}_3}\left(H\mu_{3}(N,t,z)\right)dt.
\end{split}
\label{sS1}
\end{eqnarray}

We will use the following notations:
\begin{eqnarray}
s(z)=\left( \begin{array}{cc} \tilde{a}(z) & b(z) \\ \tilde{b}(z) & a(z) \\ \end{array} \right),
\quad
S(z)=\left( \begin{array}{cc} \tilde{A}(z) & B(z) \\ \tilde{B}(z) & A(z) \\ \end{array} \right),
\quad
S_N(z)=\left( \begin{array}{cc} \tilde{A}_N(z) & B_N(z) \\ \tilde{B}_N(z) & A_N(z) \\ \end{array} \right).
\label{sm1}
\end{eqnarray}
By employing the symmetries of the spectral functions, we will show below (see section \ref{sec2.4}) that
$\left\{\tilde{a}(z),\tilde{b}(z)\right\}$, $\left\{\tilde{A}(z),\tilde{B}(z)\right\}$ and $\left\{\tilde{A}_N(z),\tilde{B}_N(z)\right\}$ can be
expressed respectively in terms of $\left\{a(z),b(z)\right\}$, $\left\{A(z),B(z)\right\}$ and $\left\{A_N(z),B_N(z)\right\}$.

Equations (\ref{sSr11})-(\ref{sSr13}), the determinant condition (\ref{det2}) and the asymptotic behavior of $\left\{\mu_j(n,t,z)\right\}_{j=1}^4$ imply
the following properties of the spectral functions:

\noindent {\it Properties of $a(z)$ and $b(z)$.}
\begin{itemize}
\item $a(z)$, $b(z)$ are analytic in the punctured complex $z$-plane $\mathbb{C}\setminus \{0\}$;
\item $a(z)$, $b(z)$, $z^{-2N}\tilde{a}(z)$, $z^{-2N}\tilde{b}(z)$ are bounded for $|z| \geq 1$;
\item $a(z)\tilde{a}(z)-b(z)\tilde{b}(z)=1$, $z\in \mathbb{C}\setminus \{0\}$;
\item \begin{eqnarray}
a(z)=\frac{1}{C(0,0)}+O(z^{-2},even),\quad b(z)=O(z^{-1},odd), \quad z\rightarrow \infty.
\label{abasp}
\end{eqnarray}
\end{itemize}

\noindent {\it Properties of $A(z)$ and $B(z)$.}
\begin{itemize}
\item $A(z)$, $B(z)$ are analytic in the punctured complex $z$-plane $\mathbb{C}\setminus \{0\}$;
\item $A(z)$, $B(z)$ are bounded in $\bar{D}_{+}$;
\item $A(z)\tilde{A}(z)-B(z)\tilde{B}(z)=1$, $z\in \mathbb{C}\setminus \{0\}$;
\item \begin{subequations}
\begin{align}
A(z)&=\hat{E}(-1,0)+O(z^{2},even),\quad B(z)=O(z,odd), ~~z\rightarrow 0,
\\
A(z)&=E(-1,T)+O(z^{-2},even),~~z\rightarrow \infty.
\end{align}
\label{ABasp}
\end{subequations}
\end{itemize}

\noindent {\it Properties of $A_N(z)$ and $B_N(z)$.}
\begin{itemize}
\item $A_N(z)$, $B_N(z)$ are analytic in the punctured complex $z$-plane $\mathbb{C}\setminus \{0\}$;
\item $A_N(z)$, $B_N(z)$ are bounded in $\bar{D}_{+}$;
\item $A_N(z)\tilde{A}_N(z)-B_N(z)\tilde{B}_N(z)=1$, $z\in \mathbb{C}\setminus \{0\}$;
\item \begin{subequations}
\begin{align}
A_N(z)&=\hat{E}(N-1,0)+O(z^{-2},even),\quad B_N(z)=O(z^{-1},odd), ~~z\rightarrow \infty,
\\
A_N(z)&=E(N-1,T)+O(z^{2},even),~~z\rightarrow 0.
\end{align}
\label{ABNasp}
\end{subequations}
\end{itemize}

In the case of integrable PDEs, it was shown in \cite{F4,F7} that the initial datum $q(x,0)$ can be expressed
in terms of the associated spectral functions through the solution of a RH problem. This problem is a {\it singular} RH problem: its solution can have poles (at possible zeros of spectral functions), therefore, residue relations have to be added to its formulation. In \cite{MFS1}, the authors present an
alternative construction of the RH problem, which is {\it regular} relative to an augmented
contour containing additional parts (see \cite{zhou2}). In such a formulation, one does not need any hypothesis on the zeros
of the associated spectral functions. Here we will extend such an approach to the construction of the RH problem 
to the discrete case.

Define the domain $D_0$ as follows:
\begin{eqnarray}
D_0=\left\{z\Big| z\in\mathbb{C},~\varepsilon<|z|<R \right\}\cup \left\{z\Big| z\in\mathbb{C},~\frac{1}{R}<|z|<\frac{1}{\varepsilon} \right\},
\label{D0}
\end{eqnarray}
where $\varepsilon$ and $R$ are chosen such that all the zeros of $a(z)$ from $|z|> 1$ are in $D_0$.
We define a sectionally holomorphic matrix-valued function $M^{(n)}(n,z)$ by:
\begin{eqnarray}
\begin{split}
M^{(n)}(n,z)&=\left\{\begin{array}{l}\frac{1}{C(n,0)}\left(\mu_3^L(n,0,z),\frac{\mu_2^R(n,0,z)}{a^*(1/z^*)}\right),
\quad z\in D_{in}\setminus \bar{D}_0,
\\
\frac{1}{C(n,0)}\left(\frac{\mu_2^L(n,0,z)}{a(z)},\mu_3^R(n,0,z)\right),
\quad z\in D_{out}\setminus \bar{D}_0,
\\
\frac{1}{C(n,0)}\left(\mu_2^L(n,0,z),\mu_2^R(n,0,z)\right), \quad z\in D_0.
\end{array}\right.
\end{split}
\label{M1}
\end{eqnarray}
Equations (\ref{asp1}) and (\ref{abasp}) imply
\begin{align}
M^{(n)}(n,z)=I+\left( \begin{array}{cc} O(z^{-2},\text{even}) & O(z,\text{odd})
\\
O(z^{-1},\text{odd}) & O(z^{2},\text{even}) \\ \end{array} \right),
\quad z\rightarrow (\infty,0).
\label{AMaspn1}
\end{align}
The limits $M^{(n)}_{\pm}(n,z)$ of $M^{(n)}(n,z)$ as $z$ approaches the oriented contour $L^{(n)}$ in the
complex $z$-plane from the corresponding side (see figure 3) are related by
\begin{eqnarray}
M^{(n)}_{-}(n,z)=M^{(n)}_{+}(n,z)J^{(n)}(n,z), \quad z\in L^{(n)},
\label{RHP1}
\end{eqnarray}
where the jump matrix $J^{(n)}(n,z)$ is defined by
\begin{eqnarray}
\begin{split}
J^{(n)}(n,z)=\left\{\begin{array}{l}
\hat{Z}^n\left( \begin{array}{cc} 1 & -\frac{b(z)}{a^*(z)} \\ \nu \frac{b^*(z)}{a(z)} & \frac{1}{|a(z)|^2} \\ \end{array} \right),
~~|z|=1,
\\
\hat{Z}^n\left( \begin{array}{cc} a^*(1/z^*) & 0 \\  \nu b^*(1/z^*) & \frac{1}{ a^*(1/z^*)} \\ \end{array} \right),
~~\{|z|=\varepsilon\}\cup \{|z|=R\},
\\
\hat{Z}^n\left( \begin{array}{cc} a(z) & -b(z) \\  0 &  \frac{1}{a(z)}\\ \end{array} \right),
~~\{|z|=\frac{1}{\varepsilon}\}\cup \{|z|=\frac{1}{R}\}.
\end{array}\right.
\end{split}
\label{JM1}
\end{eqnarray}

We define a sectionally holomorphic matrix-valued function $M^{(0)}(t,z)$ by:
\begin{eqnarray}
\begin{split}
M^{(0)}(t,z)&=\left\{\begin{array}{l}\frac{1}{\hat{E}(-1,t)}\left(\mu_1^L(0,t,z),\frac{\mu_2^R(0,t,z)}{A^*(1/z^*)}\right),
\quad z\in D_{-}\setminus \bar{D}_0,
\\
\frac{1}{\hat{E}(-1,t)}\left(\frac{\mu_2^L(0,t,z)}{A(z)},\mu_1^R(0,t,z)\right),
\quad z\in D_{+}\setminus \bar{D}_0,
\\
\frac{1}{\hat{E}(-1,t)}\left(\mu_2^L(0,t,z),\mu_2^R(0,t,z)\right), \quad z\in D_0,
\end{array}\right.
\end{split}
\label{M2}
\end{eqnarray}
where $D_0$ is of the form (\ref{D0}) containing all the zeros of $A(z)$ in $D_{+}$.
Equations (\ref{asp2}) and (\ref{ABasp}) imply
\begin{align}
M^{(0)}(t,z)=I+\left( \begin{array}{cc} O(z^{-2},\text{even}) & O(z,\text{odd})
\\
O(z^{-1},\text{odd}) & O(z^{2},\text{even}) \\ \end{array} \right),
\quad z\rightarrow (\infty,0).
\label{AMaspt1}
\end{align}
The limits $M^{(0)}_{\pm}(t,z)$ of $M^{(0)}(t,z)$ as $z$ approaches the oriented contour $L$ in the
complex $z$-plane from the corresponding side (see figure 4) are related by
\begin{eqnarray}
M^{(0)}_{-}(t,z)=M^{(0)}_{+}(t,z)J^{(0)}(t,z), \quad z\in L,
\label{RHP2}
\end{eqnarray}
where the jump matrix $J^{(0)}(t,z)$ is defined by
\begin{eqnarray}
\begin{split}
J^{(0)}(t,z)=\left\{\begin{array}{l}
e^{i\omega(z)t\hat{\sigma_3}}\left( \begin{array}{cc} 1 & -\frac{B(z)}{A^*(1/z^*)}
\\ \nu \frac{B^*(1/z^*)}{A(z)} & \frac{1}{A(z)A^*(1/z^*)} \\ \end{array} \right),
~~z\in \mathcal{L}_1,
\\
e^{i\omega(z)t\hat{\sigma_3}}\left( \begin{array}{cc} A^*(1/z^*) & 0 \\  \nu B^*(1/z^*) & \frac{1}{ A^*(1/z^*)} \\ \end{array} \right),
~~z\in \partial \{D_{-}\setminus \bar{D}_0\}\setminus \bar{\mathcal{L}}_1,
\\
e^{i\omega(z)t\hat{\sigma_3}}\left( \begin{array}{cc} A(z) & -B(z) \\  0 &  \frac{1}{A(z)}\\ \end{array} \right),
~~~~z\in \partial \{D_{+}\setminus \bar{D}_0\}\setminus \bar{\mathcal{L}}_1,
\\
I, ~~z\in \partial  (D_{+})\cap \bar{D}_0,
\end{array}\right.
\end{split}
\label{JM2}
\end{eqnarray}
with $\mathcal{L}_1=\partial  (D_{+})\setminus \bar{D}_0$.

Similarly, we define a sectionally holomorphic matrix-valued function $M^{(N)}(t,z)$:
\begin{eqnarray}
\begin{split}
M^{(N)}(t,z)&=\left\{\begin{array}{l}\frac{1}{\hat{E}(N-1,t)}\left(\mu_4^L(N,t,z),\frac{\mu_3^R(N,t,z)}{A_N^*(1/z^*)}\right),
\quad z\in D_{-}\setminus \bar{D}_0,
\\
\frac{1}{\hat{E}(N-1,t)}\left(\frac{\mu_3^L(N,t,z)}{A_N(z)},\mu_4^R(N,t,z)\right),
\quad z\in D_{+}\setminus \bar{D}_0,
\\
\frac{1}{\hat{E}(N-1,t)}\left(\mu_3^L(N,t,z),\mu_3^R(N,t,z)\right), \quad z\in D_0,
\end{array}\right.
\end{split}
\label{M3}
\end{eqnarray}
where $D_0$ is of the form (\ref{D0}) containing all the zeros of $A_N(z)$ in $D_{+}$.
Equations (\ref{asp3}) and (\ref{ABNasp}) imply
\begin{align}
M^{(N)}(t,z)=I+\left( \begin{array}{cc} O(z^{2},\text{even}) & O(z^{-1},\text{odd})
\\
O(z,\text{odd}) & O(z^{-2},\text{even}) \\ \end{array} \right),
\quad z\rightarrow (0,\infty).
\label{AMaspt2}
\end{align}
The limits $M^{(N)}_{\pm}(t,z)$ of $M^{(N)}(t,z)$ as $z$ approaches the contour $L$ are related by
\begin{eqnarray}
M^{(N)}_{-}(t,z)=M^{(N)}_{+}(t,z)J^{(N)}(t,z), \quad z\in L,
\label{RHP3}
\end{eqnarray}
where the jump matrix $J^{(N)}(t,z)$ is similar to (\ref{JM2}) with $A(z)$, $B(z)$ replaced by $A_N(z)$, $B_N(z)$.

\subsection{The symmetries}\label{sec2.4}

From the symmetry properties of $Q(n,t)$ and $H(n,t,z)$ it follows that
\begin{eqnarray}
\begin{split}
\tilde{a}(z)&=a^{\ast}(\frac{1}{z^\ast}), \quad \tilde{b}(z)=\nu b^{\ast}(\frac{1}{z^\ast}),
 \\
\tilde{A}(z)&=A^{\ast}(\frac{1}{z^\ast}), \quad \tilde{B}(z)=\nu B^{\ast}(\frac{1}{z^\ast}),
\\
\tilde{A}_N(z)&=A_N^{\ast}(\frac{1}{z^\ast}), \quad \tilde{B}_N(z,T)=\nu B_N^{\ast}(\frac{1}{z^\ast}).
\end{split}
\label{ror1}
\end{eqnarray}
Moreover, the asymptotic behavior of $\left\{\mu_j(n,t,z)\right\}_{j=1}^4$  imply the following symmetry conditions:
\begin{eqnarray}
\begin{split}
a(-z)&=a(z),\quad b(-z)=-b(z),
\\
A(-z)&=A(z),\quad B(-z)=-B(z),
\\
A_N(-z)&=A_N(z),\quad B_N(-z)=-B_N(z).
\end{split}
\label{sr}
\end{eqnarray}

\subsection{A Riemann-Hilbert problem}

We define a sectionally holomorphic matrix-valued function $M(n,t,z)$ by:
\begin{eqnarray}
\begin{split}
M(n,t,z)&=\left\{\begin{array}{l}\frac{E(N-1,t)}{C(n,t)}\left(\frac{a^*(\frac{1}{z^*})}{\alpha^*(\frac{1}{z^*})}\mu_4^L(n,t,z),\frac{\mu_2^R(n,t,z)}{a^*(\frac{1}{z^*})}\right),
\quad z\in D_{-in}\setminus \bar{D}_0,
\\
\frac{E(N-1,t)}{C(n,t)}\left(\frac{\mu_1^L(n,t,z)}{d(z)},\mu_3^R(n,t,z)\right), \quad z\in D_{-out}\setminus \bar{D}_0,
 \\
\frac{E(N-1,t)}{C(n,t)}\left(\mu_3^L(n,t,z),\frac{\mu_1^R(n,t,z)}{d^*(\frac{1}{z^*})}\right), \quad z\in D_{+in}\setminus \bar{D}_0,
\\
\frac{E(N-1,t)}{C(n,t)}\left(\frac{\mu_2^L(n,t,z)}{a(z)},\frac{a(z)}{\alpha(z)}\mu_4^R(n,t,z)\right), \quad z\in D_{+out}\setminus \bar{D}_0,
\\
\frac{E(N-1,t)}{C(n,t)}\left(\mu_2^L(n,t,z),\mu_2^R(n,t,z)\right), \quad z\in D_0,
\end{array}\right.
\end{split}
\label{M4}
\end{eqnarray}
with
\begin{eqnarray}
\begin{split}
&\alpha(z)=a(z)A_N(z)+\nu z^{-2N} b^*(\frac{1}{z^*})B_N(z),
\\
&d(z)=a(z)A^*(\frac{1}{z^*})-\nu b(z)B^*(\frac{1}{z^*}).
\end{split}
\label{grd1}
\end{eqnarray}
Here the domain $D_0$ is of the form (\ref{D0}) where $\varepsilon$ and $R$ are chosen such that
all the zeros of $a^*(\frac{1}{z^*})$, $d^*(\frac{1}{z^*})$ and $\alpha^*(\frac{1}{z^*})$ from $|z|\leq 1$ are in $D_0$.
The limits $M_{\pm}(n,t,z)$ of $M(n,t,z)$ as $z$ approaches the oriented contour $L$ in the
complex $z$-plane from the corresponding side (see figure 4) are related by:
\begin{eqnarray}
M_{-}(n,t,z)=M_{+}(n,t,z)J(n,t,z), \quad z\in L.
\label{RHP4}
\end{eqnarray}
Here the jump matrix $J(n,t,z)$ has explicit $(n,t)$-dependence
\begin{eqnarray}
J(n,t,z)=\hat{Z}^ne^{i\omega(z)t\hat{\sigma_3}}J_0(z),
\label{jm}
\end{eqnarray}
where $J_0(z)$ is defined by:
\begin{itemize}
\item
For $z\in L$, $|z|\leq 1$, we have
\begin{eqnarray}
\begin{split}
J_0=\left\{\begin{array}{l}
\left( \begin{array}{cc} 1 & -\Gamma_1(z) \\ 0 & 1 \\ \end{array} \right)
\left( \begin{array}{cc} 1 & -\gamma(z) \\ \nu \gamma^*(z) & 1-\nu |\gamma(z)|^2 \\ \end{array} \right)
\left( \begin{array}{cc} 1 & 0 \\ \nu \Gamma^*_1(z) & 1 \\ \end{array} \right),
z\in (\bar{D}_{-in}\cap \bar{D}_{+out})\setminus \bar{D}_0,
\\
\left( \begin{array}{cc} 1 & -\nu\Gamma^*(z) \\ 0 & 1 \\ \end{array} \right)
\left( \begin{array}{cc} 1-\nu |\gamma(z)|^2 & \gamma(z) \\ -\nu \gamma^*(z) & 1 \\ \end{array} \right)
\left( \begin{array}{cc} 1 & 0 \\  \Gamma(z) & 1 \\ \end{array} \right),
z\in (\bar{D}_{+in}\cap \bar{D}_{-out})\setminus \bar{D}_0,
 \\
\left( \begin{array}{cc} 1 & -\nu\Gamma^*(\frac{1}{z^*}) \\ 0 & 1 \\ \end{array} \right)
\left( \begin{array}{cc} 1 & 0 \\  \nu\Gamma_1^*(\frac{1}{z^*}) & 1 \\ \end{array} \right),
z\in (\bar{D}_{+in}\cap \bar{D}_{-in})\setminus \bar{D}_0,
\\
\left( \begin{array}{cc} a^*(\frac{1}{z^*}) & 0 \\  \nu\Gamma_2^*(\frac{1}{z^*}) & \frac{1}{ a^*(\frac{1}{z^*})} \\ \end{array} \right),
z\in \partial D_0 \cap D_{-in},
\\
\left( \begin{array}{cc} \frac{A(z)}{d^*(\frac{1}{z^*})} & -\frac{B(z)}{d^*(\frac{1}{z^*})} \\  -\nu b^*(\frac{1}{z^*}) &  a^*(\frac{1}{z^*}) \\ \end{array} \right),
z\in \partial D_0 \cap D_{+in},
\end{array}\right.
\end{split}
\label{JM4}
\end{eqnarray}
where
\begin{eqnarray}
\begin{split}
&\gamma(z)=\frac{b(z)}{a^*(\frac{1}{z^*})},
\\
&\Gamma(z)=\frac{\nu B^*(\frac{1}{z^*})/A^*(\frac{1}{z^*})}{a(z)\left(a(z)-\nu b(z) B^*(\frac{1}{z^*})/A^*(\frac{1}{z^*})\right)},
\\
&\Gamma_1(z)=\frac{z^{-2N}a(z) B_N(z)/A_N(z)}{a(z)+\nu z^{-2N} b^*(\frac{1}{z^*}) B_N(z)/A_N(z)},
\\
&\Gamma_2(z)=\frac{\Gamma_1(z)}{a(z)}+b(z).
\end{split}
\label{grd3}
\end{eqnarray}
\item
For $z\in L$, $|z|> 1$, we have
\begin{eqnarray}
J_0(z)=\left( \begin{array}{cc} -1 & 0 \\ 0 & \nu \\ \end{array} \right)
J^{\dag}(\frac{1}{z^*})
\left( \begin{array}{cc} -1 & 0 \\ 0 & \nu \\ \end{array} \right),
\label{JM5}
\end{eqnarray}
where the superscript $\dag$ denotes the Hermitian (conjugate transpose) of a matrix.
\item
For $z\in L$, $z\in D_0$, we have $J_0(z)=I$.
\end{itemize}

Equations (\ref{asp1}), and (\ref{abasp})-(\ref{ABNasp}) imply
\begin{subequations}
\begin{align}
M(n,t,z)&=I+\left( \begin{array}{cc} O(z^{-2},\text{even}) & O(z,\text{odd})
\\
O(z^{-1},\text{odd}) & O(z^{2},\text{even}) \\ \end{array} \right),
\quad z\rightarrow (\infty,0), ~~\text{for} ~~0\leq n\leq N-1;
\label{AMasp1a}
\\
M(N,t,z)&=E^2(N-1,t)I+\left( \begin{array}{cc} O(z^{2},\text{even}) & O(z^{-1},\text{odd})  \\
  O(z,\text{odd}) & O(z^{-2},\text{even}) \\ \end{array} \right),
\quad z\rightarrow (0,\infty).
\label{AMasp1b}
\end{align}
\label{AMasp1}
\end{subequations}

Substituting $\frac{C(n,t)}{E(N-1,t)}M(n,t,z)$ into (\ref{LPT3}) and making use of (\ref{AMasp1}), we find the following expression for the solution:
\begin{eqnarray}
q(n-1,t)=\lim_{z\rightarrow 0}\frac{1}{z}(M(n,t,z))^{12}.
\label{solution3}
\end{eqnarray}

\subsection {The Global relation }

We now show that the six spectral functions are not independent but satisfy an algebraic relation called global relation.
\begin{proposition}
Let the spectral functions $\{a(z), b(z)\}$, $\{A(z), B(z)\}$, $\{A_N(z), B_N(z)\}$ be defined in
equations (\ref{sm1}), where $s(z)$, $S(z)$, $S_N(z)$ are defined by equations (\ref{sSr11}), (\ref{sSr2}), (\ref{sSr3}), and $\mu_2(n,t,z)$, $\mu_3(n,t,z)$ are defined
by equations (\ref{muibv}) in terms of the smooth function $q(n, t)$. These spectral functions are
not independent but they satisfy the global relation
\begin{eqnarray}
\begin{split}
&(a(z)A_N(z)+\nu z^{-2N}b^*(\frac{1}{z^*})B_N(z))B(z)-(z^{-2N}a^*(\frac{1}{z^*})B_N(z)+b(z)A_N(z))A(z)
\\=&-e^{-2i\omega(z)T}G(z), \quad z\in\mathbb{C}\setminus\{0\},
\end{split}
\label{gr}
\end{eqnarray}
where
\begin{eqnarray*}
G(z)&=\frac{1}{C(0,T)}&\sum_{m=1}^{N} C(m,T)z^{-(2m-1)}q(m-1,T)\mu_{4}^{22}(m,T,z)).
\label{gr4}
\end{eqnarray*}
\end{proposition}

{\bf Proof} \quad Equation (\ref{s5}) implies
\begin{eqnarray}
\mu_2(n,t,z)&=&\mu_1(n,t,z)\hat{Z}^ne^{i\omega(z)t\hat{\sigma_3}}S^{-1}(z). \label{s8}
\end{eqnarray}
Evaluating (\ref{s8}) at $n=0$ and $t =T$, we find
\begin{eqnarray}
\mu_2(0,T,z)=e^{i\omega(z)T\hat{\sigma_3}}S^{-1}(z).
\label{s9}
\end{eqnarray}
Evaluating (\ref{s7}) at $n=0$ and $t =T$, we find
\begin{eqnarray}
\mu_4(0,T,z)&=&\mu_2(0,T,z)e^{i\omega(z)T\hat{\sigma_3}}\left(s(z)\hat{Z}^{-N}S_N(z)\right).
\label{s10}
\end{eqnarray}
Substituting (\ref{s9}) into (\ref{s10}), we find
\begin{eqnarray}
e^{-i\omega(z)T\hat{\sigma_3}}\mu_4(0,T,z)=S^{-1}(z)s(z)\hat{Z}^{-N} S_N(z).
\label{gr2}
\end{eqnarray}
The $(1,2)$-entry of this equation yield the global relation (\ref{gr}).  \QEDB

\section{Existence under the assumption that the global relation is valid}\label{section3}

The analysis of section \ref{section2} motivates the following definitions and results for the spectral
functions.

\begin{definition} (the spectral functions $a(z)$ and $b(z)$).
Given $q_0(n)$, define the vector $\phi(n, z) = (\phi_1(n,z), \phi_2(n,z))^\mathcal{T}$
(here and in what follows the symbol $\mathcal{T}$ denotes the transpose of a vector) as the unique solution of
\begin{eqnarray}
\begin{split}
&f_0(n)\phi_1(n+1,z)-z^2\phi_1(n,z)=zq_0(n)\phi_{2}(n,z),
\\
&f_0(n)\phi_2(n+1,z)-\phi_2(n,z)=\nu z q^*_0(n)\phi_{1}(n,z), ~0\leq n\leq N-1,~z\in \mathbb{C}\setminus \{0\},
\\
&\phi(N, z)=(0,1)^\mathcal{T},
\end{split}
\label{def1phi}
\end{eqnarray}
where $f_0(n)=\sqrt{1-\nu |q_0(n)|^2}$.
Given $\phi(n, z)$  define the spectral functions $a(z)$ and $b(z)$ by
\begin{eqnarray}
a(z)=\phi_2(0,z),~~b(z)=\phi_1(0,z), ~~z\in \mathbb{C}\setminus \{0\}.
\label{dab}
\end{eqnarray}
\end{definition}

\noindent {\it Properties of $a(z)$ and $b(z)$}:
\begin{itemize}
\item $a(z)$, $b(z)$ are analytic for $z\in\mathbb{C}\setminus \{0\}$;
\item $a(z)$, $b(z)$, $z^{-2N}a^{\ast}(\frac{1}{z^\ast})$, $z^{-2N}b^{\ast}(\frac{1}{z^\ast})$ are bounded for $|z| \geq 1$;
\item $a(-z)=a(z)$, $b(-z)=-b(z)$;
\item $a(z)a^*(\frac{1}{z^*})-\nu b(z)b^*(\frac{1}{z^*})=1$, $z\in \mathbb{C}\setminus \{0\}$;
\item $a(z)$, $b(z)$ satisfy the asymptotics
\begin{eqnarray}
a(z)=\frac{1}{C(0,0)}+O(z^{-2},even),\quad b(z)=O(z^{-1},odd), \quad z\rightarrow \infty.
\label{abasp1}
\end{eqnarray}
\end{itemize}

Definition 1 gives rise to the map
\begin{eqnarray}
\mathbb{S}: ~~\{q_0(n)\} \mapsto \left\{a(z),b(z)\right\}.
\label{map1}
\end{eqnarray}
The inverse of this map
\begin{eqnarray}
\mathbb{Q}: ~~\left\{a(z),b(z)\right\}\mapsto \{q_0(n)\},
\label{map1i}
\end{eqnarray}
can be defined as follows:
\begin{eqnarray}
q_0(n)=\lim_{z\rightarrow 0}(z^{-1}M^{(n)}(n+1,z))^{12},
\label{abis}
\end{eqnarray}
where $M^{(n)}(n, z)$ is the unique solution of the following RH problem:
\begin{itemize}
\item
$M^{(n)}(n, z)$ is a sectionally holomorphic function relative to the contour $L^{(n)}$ (see figure 3).
\item
The limits $M^{(n)}_{\pm}(n,z)$ of $M^{(n)}(n,z)$ as $z$ approaches $L^{(n)}$ are related by (\ref{RHP1}),
where the jump matrix $J^{(n)}(n,z)$ is defined in terms of the spectral functions $a(z)$ and $b(z)$ by (\ref{JM1}).
\item
The asymptotics of $M^{(n)}(n, z)$ satisfies (\ref{AMaspn1}).
\end{itemize}

\begin{definition}
Denote by $H_0(t, z)$ the matrix $H(0,t, z)$, in which $q(-1, t)$ and  $q(0, t)$ are
replaced by $g_{-1}(t)$ and  $g_{0}(t)$.
Given the smooth functions $g_{-1}(t)$ and  $g_{0}(t)$ define the vector $\varphi(t, z) = (\varphi_1(t,z), \varphi_2(t,z))^\mathcal{T}$
by the unique solution of
\begin{eqnarray}
\begin{split}
&(\varphi_1)_t-2i\omega(z)\varphi_1=H_0^{11}\varphi_1+H_0^{12}\varphi_2,
\\
&(\varphi_2)_t=H_0^{21}\varphi_1+H_0^{22}\varphi_2, ~~0<t<T, ~~z\in \mathbb{C}\setminus \{0\},
\\
&\varphi(0, z) =(0,1)^\mathcal{T}.
\end{split}
\label{ABphidef}
\end{eqnarray}
Given $\varphi(t, z)$ define the spectral functions $A(z)$ and $B(z)$ by
\begin{eqnarray}
A(z)=\varphi_2^*(T,\frac{1}{z^*}),~~B(z)=- e^{-2iw(z)T}\varphi_1(T,z), ~~z\in \mathbb{C}\setminus \{0\}.
\label{dAB1}
\end{eqnarray}
\end{definition}

\noindent {\it Properties of $A(z)$ and $B(z)$}:
\begin{itemize}
\item $A(z)$, $B(z)$ are analytic for $z\in \mathbb{C}\setminus \{0\}$;
\item $A(z)$, $B(z)$ are bounded for $z\in \bar{D}_{+}$;
\item $A(-z)=A(z)$, $B(-z)=-B(z)$;
\item $A(z)A^*(\frac{1}{z^*})-\nu B(z)B^*(\frac{1}{z^*})=1,~~ z\in \mathbb{C}\setminus\{0\}$;
\item $A(z)$, $B(z)$ satisfy the asymptotics
\begin{subequations}
\begin{align}
A(z)&=\hat{E}(-1,0)+O(z^{2},even),\quad B(z)=O(z,odd), ~~z\rightarrow 0,
\\
A(z)&=E(-1,T)+O(z^{-2},even),~~z\rightarrow \infty.
\end{align}
\label{ABasp1}
\end{subequations}
\end{itemize}

Definition 2 gives rise to the map
\begin{eqnarray}
\mathbb{S}^0: ~~\left\{g_{-1}(t),g_{0}(t)\right\} \mapsto \left\{A(z), B(z)\right\}.
\label{map2}
\end{eqnarray}
The inverse of this map
\begin{eqnarray}
\mathbb{Q}^0: ~~\left\{A(z), B(z)\right\} \mapsto \left\{g_{-1}(t),g_{0}(t)\right\},
\label{map2i}
\end{eqnarray}
can be defined as follows:
\begin{subequations}
\begin{align}
&g_{-1}(t)=\lim_{z\rightarrow 0}\left(z^{-1}M^{(0)}(t,z)\right)^{12},
\\
&g_{0}(t)=(\nu|g_{-1}(t)|^2-1)^{-1}\lim_{z\rightarrow 0}\left(i(g_{-1}(t))_t+2g_{-1}(t)-z^{-2}g_{-1}(t)(M^{(0)}(t,z))^{22}
+z^{-3}(M^{(0)}(t,z))^{12}\right),
\end{align}
\label{bvs}
\end{subequations}
where $M^{(0)}(t,z)$ is the unique solution of the following RH problem:
\begin{itemize}
\item
$M^{(0)}(t,z)$ is a sectionally holomorphic function relative to the contour $L$ (see figure 4).
\item
The limits $M^{(0)}_{\pm}(t,z)$ of $M^{(0)}(t,z)$ as $z$ approaches $L$ are related by (\ref{RHP2}),
where the jump matrix $J^{(0)}(t,z)$ is defined in terms of the spectral functions $A(z)$ and $B(z)$ by (\ref{JM2}).
\item
The asymptotics of $M^{(0)}(t,z)$ satisfies (\ref{AMaspt1}).
\end{itemize}

\begin{definition}
Denote by $H_N(t, z)$ the matrix $H(N,t, z)$, in which $q(N-1, t)$ and  $q(N, t)$ are
replaced by $g_{N-1}(t)$ and  $g_{N}(t)$.
Given the smooth functions $g_{N-1}(t)$ and  $g_{N}(t)$ define the vector $\psi(t, z) = (\psi_1(t,z), \psi_2(t,z))^\mathcal{T}$
by equations similar to (\ref{ABphidef}) with $H_0(t, z)$ replaced by $H_N(t, z)$.
Given $\psi(t, z)$ define the spectral functions $A_N(z)$ and $B_N(z)$ by
\begin{eqnarray}
A_N(z)=\psi_2^*(T,\frac{1}{z^*}),~~B_N(z)=- e^{-2iw(z)T}\psi_1(T,z), ~~z\in \mathbb{C}\setminus \{0\}.
\label{dABN1}
\end{eqnarray}
\end{definition}

\noindent {\it Properties of $A_N(z)$ and $B_N(z)$}:
\begin{itemize}
\item $A_N(z)$, $B_N(z)$ are analytic for $z\in \mathbb{C}\setminus \{0\}$;
\item $A_N(z)$, $B_N(z)$ are bounded for $z\in \bar{D}_{+}$;
\item $A_N(-z)=A_N(z)$, $B_N(-z)=-B_N(z)$;
\item $A_N(z)A_N(\frac{1}{z^*})-\nu B_N(z)B_N(\frac{1}{z^*})=1$, $z\in \mathbb{C}\setminus \{0\}$;
\item $A_N(z)$, $B_N(z)$ satisfy the asymptotics
\begin{subequations}
\begin{align}
A_N(z)&=\hat{E}(N-1,0)+O(z^{-2}),\quad B_N(z)=O(z^{-1}), ~~z\rightarrow \infty,
\\
A_N(z)&=E(N-1,T)+O(z^{2}),~~z\rightarrow 0.
\end{align}
\label{ABNasp1}
\end{subequations}
\end{itemize}

Definition 3 gives rise to the map
\begin{eqnarray}
\mathbb{S}^N: ~~\left\{g_{N-1}(t),g_{N}(t)\right\} \mapsto \left\{A_N(z), B_N(z)\right\}.
\label{map3}
\end{eqnarray}
The inverse of this map
\begin{eqnarray}
\mathbb{Q}^N: ~~\left\{A_N(z), B_N(z)\right\} \mapsto \left\{g_{N-1}(t),g_{N}(t)\right\},
\label{map3i}
\end{eqnarray}
can be defined as follows:
\begin{subequations}
\begin{align}
&g_{N}(t)=-\lim_{z\rightarrow \infty}\left(zM^{(N)}(t,z)\right)^{12},
\\
&g_{N-1}(t)=(\nu|g_{N}(t)|^2-1)^{-1}\lim_{z\rightarrow \infty}\left(i(g_{N}(t))_t-2g_{N}(t)-z^{2}g_{N}(t)(M^{(N)}(t,z))^{22}
-z^{3}(M^{(N)}(t,z))^{12}\right),
\end{align}
\label{bvNs}
\end{subequations}
where $M^{(N)}(t,z)$ is the unique solution of the following RH problem:
\begin{itemize}
\item
$M^{(N)}(t,z)$ is a sectionally holomorphic function relative to the contour $L$ (see figure 4).
\item
The limits $M^{(N)}_{\pm}(t,z)$ of $M^{(N)}(t,z)$ as $z$ approaches $L$ are related by (\ref{RHP3}),
where the jump matrix $J^{(N)}(t,z)$ is defined exactly as (\ref{JM2}) with $A(z)$, $B(z)$ replaced by $A_N(z)$, $B_N(z)$.
\item
The asymptotics of $M^{(N)}(t,z)$ satisfies (\ref{AMaspt2}).
\end{itemize}

\begin{definition}
Given $q_0(n)$ define $a(z)$, $b(z)$ according
to definition 1. Suppose that there exist smooth functions $g_{-1}(t)$, $g_{0}(t)$, $g_{N-1}(t)$, $g_{N}(t)$, such that
\begin{itemize}
\item
The associated $A(z)$, $B(z)$, $A_N(z)$, $B_N(z)$, defined according to definitions 2 and 3,
satisfy the relation
\begin{eqnarray}
\begin{split}
&(a(z)A_N(z)+\nu z^{-2N}b^*(\frac{1}{z^*})B_N(z))B(z)-(z^{-2N}a^*(\frac{1}{z^*})B_N(z)+b(z)A_N(z))A(z)
\\=&-e^{2i\omega(z)T}G(z), \quad z\in\mathbb{C}\setminus\{0\},
\end{split}
\label{grd4}
\end{eqnarray}
where $G(z)$ is analytic for $z\in\mathbb{C}\setminus\{0\}$
and $G(z)=O(\frac{1}{z})$ as $z\rightarrow \infty$, and $z^{2N}G(z)=O(z)$ as $z\rightarrow 0$.
\item
\begin{align}
g_{-1}(0)=q_0(-1), ~~g_{0}(0)=q_0(0),~~g_{N-1}(0)=q_0(N-1), ~~g_{N}(0)=q_0(N).
\label{ibvd4}
\end{align}
\end{itemize}
Then we call the functions $g_{-1}(t)$, $g_{0}(t)$, $g_{N-1}(t)$, $g_{N}(t)$ an admissible set of functions with
respect to $q_0(n)$.
\end{definition}

\begin{theorem}
Suppose that the set of functions $g_{-1}(t)$, $g_{0}(t)$, $g_{N-1}(t)$, $g_{N}(t)$, is admissible with respect to $q_0(n)$;
see definition 4. Define the spectral functions $a(z)$, $b(z)$, $A(z)$, $B(z)$, $A_N(z)$, $B_N(z)$ in terms of $q_0(n)$, $g_{-1}(t)$, $g_{0}(t)$, $g_{N-1}(t)$, $g_{N}(t)$, according to definitions 1, 2 and 3.
Define $M(n,t,z)$ as a solution of the following $2\times 2$ matrix RH problem:
\begin{itemize}
\item
$M(n,t,z)$ is sectionally holomorphic for $z\in \mathbb{C}\setminus L$; see figure 4 for the contour $L$. 
\item
For $z\in L$, $M(n,t,z)$ satisfies the jump conditions (\ref{RHP4}), where the jump matrix $J(n,t,z)$ is defined
in terms of the spectral functions $a(z)$, $b(z)$, $A(z)$, $B(z)$, $A_N(z)$, $B_N(z)$ by equations (\ref{jm})-(\ref{JM5}).
\item
The asymptotics of $M(n,t,z)$ satisfies (\ref{AMasp1}).
\end{itemize}
Then $M(n,t,z)$ exists and is unique. Define $q(n,t)$ in terms of $M(n,t,z)$ by
\begin{eqnarray}
q(n-1,t)=\lim_{z\rightarrow 0}\frac{1}{z}(M(n,t,z))^{12}.
\label{solutiontheorem}
\end{eqnarray}
Then $q(n,t)$ solves the DNLS equation (\ref{al1}) with
\begin{eqnarray}
q(n,0)=q_0(n), ~q(-1,t)=g_{-1}(t), ~q(0,t)=g_0(t),~q(N-1,t)=g_{N-1}(t), ~q(N,t)=g_N(t).
\label{ibv}
\end{eqnarray}
\end{theorem}
{\bf Proof}
The proof follows the analogous steps as in the case of integrable PDEs on the interval; see for example \cite{F7,MFS1}. The main steps are as follows.

The function $M(n, t, z)$ satisfies a regular RH problem. The unique solvability of this RH problem is
a consequence of the existence of a vanishing lemma; see \cite{zhou1,FI}.

{\it Proof that $q(n,t)$ solves the DNLS equation.}
Following the proof in the case of the DNLS equation on the non-negative integers, 
it can be verified directly that if $M(n, t, z)$ solves the above RH problem and if $q(n,t)$ is defined by (\ref{solutiontheorem}),
then $q(n,t)$ solves the DNLS equation; see \cite{XiaFokas}.

{\it Proof that $q(n,0)=q_0(n)$.} The proof that $q(n,t)$ satisfies the initial condition $q(n,0)=q_0(n)$ follows from the fact
that it is possible to map the RH problem for $M(n,0,z)$ to that for $M^{(n)}(n, z)$.
Indeed, it follows from (\ref{M1}) and (\ref{M4}) that
\begin{eqnarray}
M^{(n)}(n, z)=M(n,0,z)T^{(n)}(n,z),
\label{Mn}
\end{eqnarray}
where the transformation matrix $T^{(n)}(n,z)$ are given by
\begin{eqnarray}
\begin{split}
T^{(n)}(n,z)=\left\{\begin{array}{l}
\left( \begin{array}{cc} 1 & 0
\\ -z^{-2n}\Gamma(z) & 1 \\ \end{array} \right),
~~z\in D_{-out}\setminus \bar{D}_0,
\\
\left( \begin{array}{cc} 1 & -\nu z^{2n}\Gamma^*(\frac{1}{z^*})
\\ 0 & 1 \\ \end{array} \right),
~~z\in D_{+in}\setminus \bar{D}_0,
\\
\left( \begin{array}{cc} 1 & - z^{2n}\Gamma_1(z)
\\ 0 & 1 \\ \end{array} \right),
~~z\in D_{+out}\setminus \bar{D}_0,
\\
\left( \begin{array}{cc} 1 & 0
\\ -\nu z^{-2n}\Gamma_1^*(\frac{1}{z^*}) & 1 \\ \end{array} \right),
~~z\in D_{-in}\setminus \bar{D}_0,
\\
I,~~z\in D_0.
\end{array}\right.
\end{split}
\label{MnT}
\end{eqnarray}

{\it Proof that $q(-1,t)=g_{-1}(t)$ and $q(0,t)=g_0(t)$.} The proof that $q(n,t)$ satisfies the boundary conditions $q(-1,t)=g_{-1}(t)$ and $q(0,t)=g_0(t)$ follows from the fact
that it is possible to map the RH problem for $M(0,t,z)$ to that for $M^{(0)}(t, z)$.
In fact, it follows from (\ref{M2}) and (\ref{M4}) that
\begin{eqnarray}
M^{(0)}(t, z)=M(0,t,z)T_0^{(t)}(t,z),
\label{Mt0}
\end{eqnarray}
where the transformation matrix $T_0^{(t)}(t,z)$ are given by
\begin{eqnarray}
\begin{split}
T_0^{(t)}(t,z)=\left\{\begin{array}{l}
C(0,0)E(-1,T)\left( \begin{array}{cc} \frac{1}{d^*(\frac{1}{z^*})} & 0
\\ -\nu e^{-2i\omega(z)t}\frac{b^*(\frac{1}{z^*})}{A(z)} & d^*(\frac{1}{z^*}) \\ \end{array} \right),
~~z\in D_{+in}\setminus \bar{D}_0,
\\
C(0,0)E(-1,T)\left( \begin{array}{cc} \frac{A^*(\frac{1}{z^*})}{a^*(\frac{1}{z^*})} & 0
\\ \nu e^{-2i\omega(z)t}\frac{a^*(\frac{1}{z^*})G_1^*(\frac{1}{z^*})}{\alpha^*(\frac{1}{z^*})} & \frac{a^*(\frac{1}{z^*})}{A^*(\frac{1}{z^*})} \\ \end{array} \right),
~~z\in D_{-in}\setminus \bar{D}_0,
\\
C(0,0)E(-1,T)\left( \begin{array}{cc} \frac{a(z)}{A(z)} & e^{2i\omega(z)t}\frac{G_1(z)a(z)}{\alpha(z)}
\\ 0 & \frac{A(z)}{a(z)} \\ \end{array} \right),
~~z\in D_{+out}\setminus \bar{D}_0,
\\
C(0,0)E(-1,T)\left( \begin{array}{cc} d(z) & -e^{2i\omega(z)t}\frac{b(z)}{A^*(\frac{1}{z^*})}
\\ 0 & \frac{1}{d(z)} \\ \end{array} \right),
~~z\in D_{-out}\setminus \bar{D}_0,
\\
C(0,0)E(-1,T)I,~~z\in D_0.
\end{array}\right.
\end{split}
\label{Mt0T}
\end{eqnarray}
Here $G_1(z)$ is the left hand side of the global relation, namely,
\begin{eqnarray}
G_1(z)=(a(z)A_N(z)+\nu z^{-2N}b^*(\frac{1}{z^*})B_N(z))B(z)-(z^{-2N}a^*(\frac{1}{z^*})B_N(z)+b(z)A_N(z))A(z).
\label{G1}
\end{eqnarray}

{\it Proof that $q(N-1,t)=g_{N-1}(t)$ and $q(N,t)=g_N(t)$.}
The proof that $q(n,t)$ satisfies the boundary conditions $q(N-1,t)=g_{N-1}(t)$ and $q(N,t)=g_N(t)$ follows from the fact
that it is possible to map the RH problem for $M(N,t,z)$ to that for $M^{(N)}(t, z)$.
In this respect, the transformation matrix $T_N^{(t)}(t,z)$ is defined as follows:
\begin{eqnarray}
\begin{split}
T_N^{(t)}(t,z)=\left\{\begin{array}{l}
\frac{1}{E(N-1,t)\hat{E}(N-1,t)}\left( \begin{array}{cc} \frac{1}{A_N(z)} & -e^{2i\omega(z)t}\frac{z^{2N}G_1(z)}{d^*(\frac{1}{z^*})}
\\ 0 & A_N(z) \\ \end{array} \right),
~~z\in D_{+in}\setminus \bar{D}_0,
\\
\frac{1}{E(N-1,t)\hat{E}(N-1,t)}\left( \begin{array}{cc} \frac{\alpha^*(\frac{1}{z^*})}{a^*(\frac{1}{z^*})} & e^{2i\omega(z)t}\frac{z^{2N}b(z)}{a^*(\frac{1}{z^*})A_N^*(\frac{1}{z^*})}
\\ 0 & \frac{a^*(\frac{1}{z^*})}{\alpha^*(\frac{1}{z^*})} \\ \end{array} \right),
~~z\in D_{-in}\setminus \bar{D}_0,
\\
\frac{1}{E(N-1,t)\hat{E}(N-1,t)}\left( \begin{array}{cc} \frac{a(z)}{\alpha(z)} & 0
\\ \nu e^{-2i\omega(z)t}\frac{z^{-2N}b^*(\frac{1}{z^*})}{a(z)A_N(z)} & \frac{\alpha(z)}{a(z)} \\ \end{array} \right),
~~z\in D_{+out}\setminus \bar{D}_0,
\\
\frac{1}{E(N-1,t)\hat{E}(N-1,t)}\left( \begin{array}{cc} A^*_N(\frac{1}{z^*}) & 0
\\ -\nu e^{-2i\omega(z)t}\frac{z^{-2N}G^*_1(\frac{1}{z^*})}{d(z)} & \frac{1}{A^*_N(\frac{1}{z^*})} \\ \end{array} \right),
~~z\in D_{-out}\setminus \bar{D}_0,
\\
\frac{1}{E(N-1,t)\hat{E}(N-1,t)}\left( \begin{array}{cc} a^*(\frac{1}{z^*}) & e^{2i\omega(z)t}z^{2N}b(z)
\\ \nu e^{-2i\omega(z)t}z^{-2N}b^*(\frac{1}{z^*}) & a(z) \\ \end{array} \right),
~~z\in D_0.
\end{array}\right.
\end{split}
\label{MtNT}
\end{eqnarray}

\section{Elimination of the unknown boundary values}\label{section4}

In this section, we will show how to characterize the unknown boundary values $g_{0}(t)$ and $g_{N-1}(t)$ in terms of the given initial value $q_0(n)$ and boundary
values $g_{-1}(t)$ and $g_{N}(t)$.
The approach we employ is the discrete analogue of the approach presented in \cite{FL1,FL2,FL3}, which uses certain asymptotic considerations of the global relation.

We introduce the functions $\left\{\phi_1(t,z), \phi_2(t,z)\right\}$ and $\left\{\varphi_1(t,z), \varphi_2(t,z)\right\}$ as follows
\begin{eqnarray}
\mu_2(0,t,z)&=&\left( \begin{array}{cc} \phi_2^*(t,\frac{1}{z^*}) & \phi_1(t,z) \\  \nu \phi_1^*(t,\frac{1}{z^*})  & \phi_2(t,z) \\ \end{array} \right),
\label{mu20}
\\
\mu_3(N,t,z)&=&\left( \begin{array}{cc} \varphi_2^*(t,\frac{1}{z^*}) & \varphi_1(t,z) \\  \nu \varphi_1^*(t,\frac{1}{z^*})  & \varphi_2(t,z) \\ \end{array} \right),
\label{mu3N}
\end{eqnarray}
Then $\phi_1(t,z)$ and $\phi_2(t,z)$ satisfy the following system of nonlinear integral equations:
\begin{eqnarray}
\begin{split}
\phi_1(t,z)=&i\int_0^te^{2i\omega(z)(t-t')}[\left( zg_{0}(t')-z^{-1}g_{-1}(t')\right)\phi_2 (t',z)-\nu\text{Re}\left( g_{0}(t')g^*_{-1}(t')\right)\phi_1 (t',z)]dt',
\\
\phi_2(t,z)=&1+i\nu\int_0^t[\left( zg^*_{-1}(t')- z^{-1}g^*_{0}(t')\right)\phi_1 (t',z)+\text{Re}\left( g_{0}(t')g^*_{-1}(t')\right)\phi_2 (t',z)]dt';
\end{split}
\label{phi}
\end{eqnarray}
whereas $\varphi_1(t,z)$ and $\varphi_2(t,z)$ satisfy the equations:
\begin{eqnarray}
\begin{split}
\varphi_1(t,z)=&i\int_0^te^{2i\omega(z)(t-t')}[\left( zg_{N}(t')-z^{-1}g_{N-1}(t')\right)\varphi_2 (t',z)
\\&\qquad-\nu\text{Re}\left( g_{N}(t')g^*_{N-1}(t')\right)\varphi_1 (t',z)]dt',
\\
\varphi_2(t,z)=&1+i\nu\int_0^t[\left( zg^*_{N-1}(t')- z^{-1}g^*_{N}(t')\right)\varphi_1 (t',z)
\\&\qquad+\text{Re}\left( g_{N}(t')g^*_{N-1}(t')\right)\varphi_2 (t',z)]dt'.
\end{split}
\label{vphi}
\end{eqnarray}

Equation (\ref{sSr2}) implies
\begin{eqnarray}
A(z)=\phi_2^*(T,\frac{1}{z^*}),\quad B(z)=-e^{-2i\omega(z)T} \phi_1(T,z).
\label{ABvp}
\end{eqnarray}
Equation (\ref{sSr3}) implies
\begin{eqnarray}
A_N(z)=\varphi_2^*(T,\frac{1}{z^*}),\quad B_N(z)=-e^{-2i\omega(z)T} \varphi_1(T,z).
\label{ABNvp}
\end{eqnarray}
Following (\ref{asp2}) we find
\begin{eqnarray}
\begin{split}
\phi_1(t,z)&=\left(E(-1,t)g_{-1}(t)-e^{2i\omega(z)t} g_{-1}(0)\right)z+O(z^{3}), \quad z\rightarrow 0, \quad z\in D_{-in},
\\
\phi_2(t,z)&=E(-1,t)+O(z^{2}), \quad z\rightarrow 0, \quad z\in D_{-in}.
\end{split}
\label{aspvp1}
\end{eqnarray}
Following (\ref{asp3}) we find
\begin{eqnarray}
\begin{split}
\varphi_1(t,z)&=-\left(E(N-1,t)g_N(t)-e^{2i\omega(z)t} g_N(0)\right)\frac{1}{z}+O(z^{-3}), \quad z\rightarrow \infty, \quad z\in D_{-out},
\\
\varphi_2(t,z)&=E(N-1,t)+O(z^{-2}), \quad z\rightarrow \infty, \quad z\in D_{-out}.
\end{split}
\label{aspvp2}
\end{eqnarray}

We now express the global relation in terms of the eigenfunctions $\left\{\phi_j(t,z)\right\}_{j=1}^2$, $\left\{\varphi_j(t,z)\right\}_{j=1}^2$.
For each $t \in (0, T)$, let $R(n, t, z)$ be the solution of the $n$-part of the Lax pair
of (\ref{LPS3}) such that $R(N, t, z)=I$, i.e., $R(n, t, z)$ is the unique solution of the equation
\begin{eqnarray}
R(n, t, z)=\frac{1}{C(n,t)}\left(I-\sum_{m=n+1}^{N}C(m,t)\hat{Z}^{-(m-n-1)}(Q(m-1,t)R(m,t,z))Z\right).
\label{R}
\end{eqnarray}
Then $R(n, t, z)$ is related to $\mu_3(n, t, z)$ by
\begin{eqnarray}
R(n,t,z)=\mu_3(n,t,z)\hat{Z}^{n-N}\mu_3^{-1}(N,t,z).
\label{Rmu1}
\end{eqnarray}
Substituting (\ref{s4}) into (\ref{Rmu1}), we find
\begin{eqnarray}
R(n,t,z)=\mu_2(n,t,z)\hat{Z}^{n}e^{i\omega(z)t\hat{\sigma_3}}\left(s(z)\right)\hat{Z}^{n-N}\mu_3^{-1}(N,t,z).
\label{Rmu2}
\end{eqnarray}
Evaluating (\ref{Rmu2}) at $n=0$, we find
\begin{eqnarray}
R(0,t,z)=\mu_2(0,t,z)e^{i\omega(z)t\hat{\sigma_3}}\left(s(z)\right)\hat{Z}^{-N}\mu_3^{-1}(N,t,z).
\label{Rmu3}
\end{eqnarray}
Using (\ref{mu20}) and (\ref{mu3N}), the $(1,2)$-entry of equation (\ref{Rmu3}) yields the global relation
\begin{eqnarray}
\begin{split}
&\phi_1(t,z)\left(a(z)\varphi_2^*(t,\frac{1}{z^*})-\nu z^{-2N}e^{-2i\omega(z)t} b^{\ast}(\frac{1}{z^\ast})\varphi_1(t,z)\right)
\\&+\phi_2^*(t,\frac{1}{z^*})(e^{2i\omega(z)t}b(z)\varphi_2^*(t,\frac{1}{z^*})-z^{-2N}a^*(\frac{1}{z^*})\varphi_1(t,z))=G(t,z), \quad z\in\mathbb{C}\setminus\{0\},
\end{split}
\label{grvp}
\end{eqnarray}
where $G(t,z)=R^{12}(0,t,z)$ is the $(1,2)$-entry of $R(0,t,z)$, and $R(n,t,z)$ is defined by equation (\ref{R}).
From (\ref{R}), we find
\begin{eqnarray}
\begin{split}
G(t,z)&=-\frac{1}{C(0,t)}\left(g_0(t)z^{-1}+\cdots+g_{N-1}(t)z^{-(2N-1)}\right)
\\&=-\frac{1}{C(0,0)E(-1,t)E(N-1,t)}\left(g_0(t)z^{-1}+\cdots+g_{N-1}(t)z^{-(2N-1)}\right),
\end{split}
\label{gasp}
\end{eqnarray}
where we have used (\ref{ct1}).

For simplicity, we shall consider here the case of zero initial conditions: $q_{0}(n)\equiv0$, which yields $C(n,0)\equiv 1$ and $a(z) \equiv 1$, $b(z)\equiv 0$.
 In this case, the global relation (\ref{grvp}) reduces to
\begin{eqnarray}
\begin{split}
&\phi_1(t,z)\varphi_2^*(t,\frac{1}{z^*})-z^{-2N}\varphi_1(t,z))\phi_2^*(t,\frac{1}{z^*})=G(t,z), \quad z\in\mathbb{C}\setminus\{0\},
\end{split}
\label{zgrvp}
\end{eqnarray}
and (\ref{gasp}) becomes
\begin{eqnarray}
\begin{split}
G(t,z)=-\frac{1}{E(-1,t)E(N-1,t)}\left(g_0(t)z^{-1}+\cdots+g_{N-1}(t)z^{-(2N-1)}\right).
\end{split}
\label{zgasp}
\end{eqnarray}

\begin{proposition}
Let $q_{0}(n)\equiv0$ for $n\in\mathbb{N}$. Let $\partial D_{\pm in}$ and $\partial D_{\pm out}$ denote the boundaries of the domains $D_{\pm in}$ and $D_{\pm out}$ defined in (\ref{D1}), oriented
so that $D_{\pm in}$ and $D_{\pm out}$ lie in the left of $\partial D_{\pm in}$ and $\partial D_{\pm out}$. Let $\left\{\phi_j(t,z)\right\}_{j=1}^2$ and $\left\{\varphi_j(t,z)\right\}_{j=1}^2$ be defined by (\ref{mu20}) and (\ref{mu3N}).
The spectral functions $A(z)$ and $B(z)$ are given by (\ref{ABvp})
where $\phi_1(t,z)$ and $\phi_2(t,z)$ satisfy the system of integral equations (\ref{phi}).
The spectral functions $A_N(z)$ and $B_N(z)$ are given by (\ref{ABNvp}) where $\varphi_1(t,z)$ and $\varphi_2(t,z)$ satisfy the system of integral equations (\ref{vphi}).
The unknown boundary values $g_{0}(t)$ and $g_{N-1}(t)$ associated with the DNLS equation (\ref{al1}) are given by the following expressions:
\begin{eqnarray}
\begin{split}
g_{0}(t)&=-g_{-1}(t)+\frac{1}{\pi i}\int_{\partial D_{-in}}\frac{1}{z^2}\chi_1 (t,z)dz
\\&+\frac{E(-1,t)E(N-1,t)}{\pi i}\int_{\partial D_{-in}}
\frac{1}{z^2}\left(\phi_1(t,\frac{1}{z})(1-\varphi_2^*(t,z^{*}))+z^{2N}\varphi_1(t,\frac{1}{z})(\phi_2^*(t,z^{*})-1)\right)dz,
\end{split}
\label{DN1}
\end{eqnarray}
\begin{eqnarray}
\begin{split}
g_{N-1}(t)&=-g_N(t)+\frac{1}{\pi i}\int_{\partial D_{-in}}\frac{1}{z^2}\hat{\chi}_1 (t,z)dz
\\&-\frac{E(-1,t)E(N-1,t)}{\pi i}\int_{\partial D_{-in}}
\frac{1}{z^2}\left(\varphi_1(t,z)(1-\phi_2^*(t,\frac{1}{z^{*}}))+z^{2N}\phi_1(t,z)(\varphi_2^*(t,\frac{1}{z^{*}})-1)\right)dz,
\end{split}
\label{DN2}
\end{eqnarray}
where
\begin{eqnarray}
\chi_j(t,z)&=&\frac{1}{E(-1,t)}\phi_j(t,z)-E(-1,t)E(N-1,t)\phi_j(t,z^{-1}),\quad j=1,2, \label{chinls}
\\
\hat{\chi}_j(t,z)&=&E(-1,t)E(N-1,t)\varphi_j(t,z)-\frac{1}{E(N-1,t)}\varphi_j(t,z^{-1}),\quad j=1,2.
\label{hchinls}
\end{eqnarray}
\end{proposition}
{\bf Proof} \quad Using (\ref{chinls}) we find
\begin{eqnarray}
\begin{split}
\int_{\partial D_{-in}}\frac{1}{z^2}\chi_1 (t,z)dz=&\frac{1}{E(-1,t)}\int_{\partial D_{-in}}\frac{1}{z^2}\phi_1(t,z)dz
\\&-E(-1,t)E(N-1,t)\int_{\partial D_{-in}}\frac{1}{z^2}\phi_1(t,\frac{1}{z})dz.
\end{split}
\label{psi2dnlsin}
\end{eqnarray}
Using the asymptotics of $\phi_1(t,z)$ (the first of equations (\ref{aspvp1})),
the first integral in the right hand side of (\ref{psi2dnlsin}) can be evaluated explicitly:
\begin{eqnarray}
\frac{1}{E(-1,t)}\int_{\partial D_{-in}}\frac{1}{z^2}\phi_1(t,z)dz=g_{-1}(t)\pi i,
\label{psi2dnlsin1}
\end{eqnarray}
where we have used the fact that the term involving $q(-1,0)$ in (\ref{aspvp1}) vanishes in (\ref{psi2dnlsin1})
due to the term $e^{2i\omega(z)t}$ which is bounded and decays as $z\rightarrow 0$ in $D_{-in}$.
Replacing in the second integral in the right hand side of (\ref{psi2dnlsin}) $z^{-1}$ by $z$, we find
\begin{eqnarray}
 -E(-1,t)E(N-1,t)\int_{\partial D_{-in}}\frac{1}{z^2}\phi_1(t,\frac{1}{z})dz=E(-1,t)E(N-1,t)\int_{\partial D_{-out}}\phi_1(t,z)dz.
\label{psi2dnlsin2}
\end{eqnarray}
Employing the global relation (\ref{zgrvp}) and the equation (\ref{zgasp}), the integral (\ref{psi2dnlsin2}) can be evaluated as follows:
\begin{eqnarray}
\begin{split}
&E(-1,t)E(N-1,t)\int_{\partial D_{-out}}\phi_1(t,z)dz
\\=&E(-1,t)E(N-1,t)\left(\int_{\partial D_{-out}}(\phi_1(t,z)-G(t,z))dz+\int_{\partial D_{-out}}G(t,z)dz\right)
\\=&E(-1,t)E(N-1,t)\int_{\partial D_{-out}}\left(\phi_1(t,z)(1-\varphi_2^*(t,\frac{1}{z^*}))+z^{-2N}\varphi_1(t,z)\phi_2^*(t,\frac{1}{z^*})\right)dz+g_0(t)\pi i
\\=&g_0(t)\pi i
\\&+E(-1,t)E(N-1,t)\int_{\partial D_{-out}}\left(\phi_1(t,z)(1-\varphi_2^*(t,\frac{1}{z^*}))+z^{-2N}\varphi_1(t,z)(\phi_2^*(t,\frac{1}{z^*})-1)\right)dz
\\=&g_0(t)\pi i
\\&-E(-1,t)E(N-1,t)\int_{\partial D_{-in}}
\frac{1}{z^2}\left(\phi_1(t,\frac{1}{z})(1-\varphi_2^*(t,z^{*}))+z^{2N}\varphi_1(t,\frac{1}{z})(\phi_2^*(t,z^{*})-1)\right)dz.
\end{split}
\label{psi2dnlsin3}
\end{eqnarray}
Equations (\ref{psi2dnlsin})-(\ref{psi2dnlsin3}) imply the formula (\ref{DN1}).

Using (\ref{hchinls}) we find
\begin{eqnarray}
\begin{split}
\int_{\partial D_{-in}}\frac{1}{z^2}\hat{\chi}_1 (t,z)dz=&E(-1,t)E(N-1,t)\int_{\partial D_{-in}}\frac{1}{z^2}\varphi_1(t,z)dz
\\&-\frac{1}{E(N-1,t)}\int_{\partial D_{-in}}\frac{1}{z^2}\varphi_1(t,\frac{1}{z})dz.
\end{split}
\label{vpsi2dnlsin}
\end{eqnarray}
Employing the global relation (\ref{zgrvp}) and the equation (\ref{zgasp}), the first integral in the right hand side of (\ref{vpsi2dnlsin}) can be evaluated as follows:
\begin{eqnarray}
\begin{split}
&E(-1,t)E(N-1,t)\int_{\partial D_{-in}}\frac{1}{z^2}\varphi_1(t,z)dz
\\=&E(-1,t)E(N-1,t)\left(\int_{\partial D_{-in}}\frac{1}{z^2}(\varphi_1(t,z)+z^{2N}G(t,z))dz-\int_{\partial D_{-in}}z^{2(N-1)}G(t,z)dz\right)
\\=&g_{N-1}(t)\pi i
\\&+E(-1,t)E(N-1,t)\int_{\partial D_{-in}}\frac{1}{z^2}\left(\varphi_1(t,z)(1-\phi_2^*(t,\frac{1}{z^*}))+z^{2N}\phi_1(t,z)\varphi_2^*(t,\frac{1}{z^*})\right)dz
\\=&g_{N-1}(t)\pi i
\\&+E(-1,t)E(N-1,t)\int_{\partial D_{-in}}\frac{1}{z^2}\left(\varphi_1(t,z)(1-\phi_2^*(t,\frac{1}{z^*}))+z^{2N}\phi_1(t,z)(\varphi_2^*(t,\frac{1}{z^*})-1)\right)dz.
\end{split}
\label{vpsi2dnlsin3}
\end{eqnarray}
Using the asymptotics of $\varphi_1(t,z)$ (the first of equations (\ref{aspvp2})),
the second integral in the right hand side of (\ref{vpsi2dnlsin}) can be evaluated explicitly:
\begin{eqnarray}
-\frac{1}{E(N-1,t)}\int_{\partial D_{-in}}\frac{1}{z^2}\varphi_1(t,\frac{1}{z})dz=\frac{1}{E(N-1,t)}\int_{\partial D_{-out}}\varphi_1(t,z)dz=g_N(t)\pi i.
\label{vpsi2dnlsin1}
\end{eqnarray}
Equations (\ref{vpsi2dnlsin})-(\ref{vpsi2dnlsin1}) imply the formula (\ref{DN2}).
\QEDB

In what follows we describe an effective characterization of the unknown boundary value $g_{0}(t)$ and $g_{N-1}(t)$ by employing a suitable perturbation expansion.
We consider  the expansions of $\phi_j(t,z)$, $\varphi_j(t,z)$, $q(m,t)$ in the following forms:
\begin{eqnarray}
\begin{split}
\phi_j(t,z)&=\phi_{j,0}(t,z)+\phi_{j,1}(t,z)\epsilon+\phi_{j,2}(t,z)\epsilon^2+\cdots, \quad j=1,2,
\\
\varphi_j(t,z)&=\varphi_{j,0}(t,z)+\varphi_{j,1}(t,z)\epsilon+\varphi_{j,2}(t,z)\epsilon^2+\cdots, \quad j=1,2,
\\
g_m(t)&=g_{m,1}(t)\epsilon+g_{m,2}(t)\epsilon^2+\cdots, \quad m=-1,0,N-1,N,
\end{split}
\label{expvp}
\end{eqnarray}
where $\epsilon$ is a small perturbation parameter.
Substituting (\ref{expvp}) into (\ref{phi}) we find
\begin{eqnarray}
\begin{split}
\phi_{1,0}(t,z)=&0,\quad \phi_{2,0}(t,z)=1, \quad \phi_{2,1}(n,z)=0,
\\
\phi_{1,1}(t,z)=&i\int_0^te^{2i\omega(z)(t-t')}\left( zg_{0,1}(t')-z^{-1}g_{-1,1}(t')\right)dt',
\\
\phi_{1,2}(t,z)=&i\int_0^te^{2i\omega(z)(t-t')}\left( zg_{0,2}(t')-z^{-1}g_{-1,2}(t')\right)dt',
\\
\phi_{2,2}(t,z)=&i\nu\int_0^t[\left( zg^*_{-1,1}(t')- z^{-1}g^*_{0,1}(t')\right)\phi_{1,1} (t',z)
+\text{Re}\left( g_{0,1}(t')g^*_{-1,1}(t')\right)]dt',
\\
\phi_{1,3}(t,z)=&i\int_0^te^{2i\omega(z)(t-t')}[ zg_{0,3}(t')-z^{-1}g_{-1,3}(t')+(zg_{0,1}(t')-z^{-1}g_{-1,1}(t'))\phi_{2,2}(t',z)
\\&\qquad -\nu\text{Re}\left( g_{0,1}(t')g^*_{-1,1}(t')\right)\phi_{1,1}(t',z)]dt',
\\
\phi_{2,3}(t,z)=&i\nu\int_0^t[\left( zg^*_{-1,1}(t')- z^{-1}g^*_{0,1}(t')\right)\phi_{1,2} (t',z)+\left( zg^*_{-1,2}(t')- z^{-1}g^*_{0,2}(t'))\right)\phi_{1,1} (t',z)
\\&+\text{Re}\left( g_{0,1}(t')g^*_{-1,2}(t')+ g_{-1,1}(t')g^*_{0,2}(t')\right)]dt'.
 \end{split}
\label{expp1}
\end{eqnarray}
The formulae for $\varphi_{j,k}(t,z)$ are similar to those of $\phi_{j,k}(t,z)$ where $g_{0,k}(t')$ and $g_{-1,k}(t')$ are replaced by $g_{N,k}(t')$ and $g_{N-1,k}(t')$, respectively.

Equation (\ref{nota1}) implies the expansion
\begin{eqnarray}
\begin{split}
E(-1,t)&=1+E_2(-1,t)\epsilon^2+E_3(-1,t)\epsilon^3+O(\epsilon^4),
\\
E(N-1,t)&=1+E_2(N-1,t)\epsilon^2+E_3(N-1,t)\epsilon^3+O(\epsilon^4),
\end{split}
\label{ct1exp}
\end{eqnarray}
where
\begin{eqnarray}
\begin{split}
E_2(-1,t)&=-\nu \int_0^t\text{Im}\left(g_{0,1}(t')g^*_{-1,1}(t')\right)dt',
\\
E_3(-1,t)&=-\nu \int_0^t\text{Im}\left(g_{0,1}(t')g^*_{-1,2}(t')+g_{0,2}(t')g^*_{-1,1}(t')\right)dt',
\\
E_2(N-1,t)&=\nu \int_0^t\text{Im}\left(g_{N,1}(t')g^*_{N-1,1}(t'))\right)dt',
\\
E_3(N-1,t)&=\nu \int_0^t\text{Im}\left(g_{N,1}(t')g^*_{N-1,2}(t')+g_{N,2}(t')g^*_{N-1,1}(t')\right)dt'.
\label{ct1expc}
\end{split}
\end{eqnarray}
Using equations (\ref{chinls}), (\ref{hchinls}) and (\ref{ct1exp}) we find
\begin{eqnarray}
\begin{split}
\chi_1 (t,z)=\chi_{1,1} (t,z)\epsilon+\chi_{1,3} (t,z)\epsilon^3+O(\epsilon^4),
\\
\hat{\chi}_1 (t,z)=\hat{\chi}_{1,1} (t,z)\epsilon+\hat{\chi}_{1,3} (t,z)\epsilon^3+O(\epsilon^4),
\end{split}
\label{psinlsexp}
\end{eqnarray}
where
\begin{eqnarray}
\begin{split}
\chi_{1,1} (t,z)=&\phi_{1,1}(t,z)-\phi_{1,1}(t,\frac{1}{z}),
\\
\chi_{1,3} (t,z)=&\phi_{1,3}(t,z)-\phi_{1,3}(t,\frac{1}{z})-E_2(-1,t)\phi_{1,1}(t,z)-(E_2(-1,t)+E_2(N-1,t))\phi_{1,1}(t,\frac{1}{z}),
\\
\hat{\chi}_{1,1} (t,z)=&\varphi_{1,1}(t,z)-\varphi_{1,1}(t,\frac{1}{z}),
\\
\hat{\chi}_{1,3} (t,z)=&\varphi_{1,3}(t,z)-\varphi_{1,3}(t,\frac{1}{z})+E_2(N-1,t)\varphi_{1,1}(t,\frac{1}{z})+(E_2(-1,t)+E_2(N-1,t))\varphi_{1,1}(t,z).
\label{psinlsexpc}
\end{split}
\end{eqnarray}
Substituting expansions (\ref{expvp}), (\ref{ct1exp}) and (\ref{psinlsexp}) into (\ref{DN1}),
we can obtain the expansions for $g_{0}(t)$ and $g_{N-1}(t)$ to all orders of $\epsilon$.
For example, the first few terms are given by the following formulae:
\begin{eqnarray}
\begin{split}
g_{0,1}(t)=&-g_{-1,1}(t)+\frac{1}{\pi i}\int_{\partial D_{-in}}\frac{1}{z^2}\left(\phi_{1,1}(t,z)-\phi_{1,1}(t,\frac{1}{z})\right)dz,
\\
g_{0,2}(t)=&-g_{-1,2}(t),
\\
g_{0,3}(t)=&-g_{-1,3}(t)+\frac{1}{\pi i}\int_{\partial D_{-in}}\frac{1}{z^2}\chi_{1,3} (t,z)dz
\\ \qquad&+\frac{1}{\pi i}\int_{\partial D_{-in}}
\frac{1}{z^2}(z^{2N}\varphi_{1,1}(t,\frac{1}{z})\phi^*_{2,2}(t,z^*)-\phi_{1,1}(t,\frac{1}{z})\varphi^*_{2,2}(t,z^*))dz,
\\
g_{N-1,1}(t)=&-g_{N,1}(t)+\frac{1}{\pi i}\int_{\partial D_{-in}}\frac{1}{z^2}\left(\varphi_{1,1}(t,z)-\varphi_{1,1}(t,\frac{1}{z})\right)dz,
\\
g_{N-1,2}(t)=&-g_{N,2}(t),
\\
g_{N-1,3}(t)=&-g_{N,3}(t)+\frac{1}{\pi i}\int_{\partial D_{-in}}\frac{1}{z^2}\hat{\chi}_{1,3} (t,z)dz
\\ \qquad&-\frac{1}{\pi i}\int_{\partial D_{-in}}
\frac{1}{z^2}(z^{2N}\phi_{1,1}(t,z)\varphi^*_{2,2}(t,\frac{1}{z^*})-\varphi_{1,1}(t,z)\phi^*_{2,2}(t,\frac{1}{z^*}))dz.
\label{DN1exp}
\end{split}
\end{eqnarray}

\section {The linearizable boundary conditions} \label{section5}

It was shown in section \ref{section4} that the spectral functions $\{A(z), B(z)\}$ and $\{A_N(z), B_N(z)\}$
can be expressed in terms of the given boundary data through the solution of a system of nonlinear integral equations.
In general it is not possible to solve directly the spectral functions $\{A(z), B(z)\}$ and $\{A_N(z), B_N(z)\}$
only in terms of the known data.
However, there exists a particular class of boundary conditions, called linearizable,
for which one can compute explicitly all spectral data necessary to construct the RH problem from
the initial data. We will identify such linearizable boundary conditions by using the algebraic manipulation of
the global relation \cite{F3,F4}.

In the case $T=\infty$, the global relation (\ref{gr}) becomes
\begin{eqnarray}
\begin{split}
(a(z)A_N(z)+\nu z^{-2N}b^*(\frac{1}{z^*})B_N(z))B(z)-(z^{-2N}a^*(\frac{1}{z^*})B_N(z)+b(z)A_N(z))A(z)=0, ~z\in \bar{D}_{+}.
\end{split}
\label{grsim1}
\end{eqnarray}
Note that the dependence of the jump matrices (\ref{jm})-(\ref{JM5}) on the boundary values is expressed in terms of the ratios $B(z)/A(z)$ and $B_N(z)/A_N(z)$.
Let
\begin{eqnarray}
\rho(z)=\frac{B(z)}{A(z)}, ~~\rho_1(z)=\frac{B_N(z)}{A_N(z)}.
\label{rho}
\end{eqnarray}
Equation (\ref{grsim1}) can be  formulated into
\begin{eqnarray}
\begin{split}
a(z)\rho(z)+\nu z^{-2N}b^*(\frac{1}{z^*})\rho(z)\rho_1(z)-z^{-2N}a^*(\frac{1}{z^*})\rho_1(z)-b(z)=0,
\quad z\in \bar{D}_{+}.
\end{split}
\label{grsim2}
\end{eqnarray}

In order to compute $\rho(z)$ and $\rho_1(z)$,
we need to derive an additional symmetry for spectral functions $\{A(z), B(z)\}$ and $\{A_N(z), B_N(z)\}$.
Following \cite{F6} and \cite{S}, we find the following result.
\begin{proposition}
Let $z\mapsto f(z)$ be the transformation in the complex $z$-plane which leaves
$\omega(z)$ invariant, i.e., $\omega(f(z))=\omega(z)$, $f(z)\neq z$.
Let $V_0(t,z)=i\omega(z)\sigma_3+H(0,t,z)$.
If there exists a $t$-independent non-singular matrix $P_0(z)$ such that
\begin{eqnarray}
V_0(t,f(z))P_0(z)=P_0(z)V_0(t,z),
\label{NV1}
\end{eqnarray}
then the spectral functions $\{A(z), B(z)\}$ possess the following symmetry
properties:
\begin{eqnarray}
\begin{split}
A(f(z))\det P_0(z)&=P_0^{22}(z)\left(P_0^{11}(z)A(z)+P_0^{12}(z)B(z)\right)
\\&~~~~-P_0^{21}(z)e^{2i\omega(z)T}\left(P_0^{12}(z)A^{\ast}(\frac{1}{z^\ast})+\nu P_0^{11}(z) B^{\ast}(\frac{1}{z^\ast})\right),
\\
B(f(z))\det P_0(z)&=P_0^{22}(z)\left(P_0^{21}(z)A(z)+P_0^{22}(z)B(z)\right)
\\&~~~~-P_0^{21}(z)e^{2i\omega(z)T}\left(P_0^{22}(z)A^{\ast}(\frac{1}{z^\ast})+\nu P_0^{21}(z)B^{\ast}(\frac{1}{z^\ast})\right).
\end{split}
\label{AB}
\end{eqnarray}
Similarly, let $V_N(t,z)= i\omega(z)\sigma_3+H(N,t,z)$. If there exists a matrix $P_N(z)$
such that $V_N(t,z)$ and $P_N(z)$ satisfy a relation in form of (\ref{NV1}),
then the spectral functions $\{A_N(z), B_N(z)\}$ possess the symmetry in the form of (\ref{AB}),
where $P_0(z)$ is replaced by $P_N(z)$.
\end{proposition}

A necessary condition for the existence of $P_0(z)$ is that
$\det V_0(t,f(z))=\det V_0(t,z)$, which in turn yields
\begin{eqnarray}
(z^{-2}-f^{-2}(z))g_0(t)g_{-1}^*(t)+(z^{2}-f^{2}(z))g_0^*(t)g_{-1}(t)=0.
\label{LC}
\end{eqnarray}
The invariance of $\omega(z)$ yields $f(z)=\frac{1}{z}$.
Then equation (\ref{LC}) becomes
\begin{eqnarray}
g_0(t)g_{-1}^*(t)-g_0^*(t)g_{-1}(t)=0.
\label{LCDNLS}
\end{eqnarray}
There are three particular solutions of the above equation:
\begin{eqnarray}
g_0(t)=0, ~~g_{-1}(t)=0, ~~g_{-1}(t)=c_1g_0(t),
\label{LCDNLSs}
\end{eqnarray}
where $c_1$ is an arbitrary real constant.
With an analogous argument for $g_{N-1}(t)$ and $g_{N}(t)$,
we find the linearizable boundary conditions are
\begin{eqnarray}
g_N(t)=0, ~~g_{N-1}(t)=0, ~~g_{N-1}(t)=c_2g_N(t),
\label{LCDNLSs}
\end{eqnarray}
where $c_2$ is an arbitrary real constant.

For economy of presentation, here we only consider the case of $g_{-1}(t)=0$ and $g_N(t)=0$; other cases can be discussed in a very similar manner.
In this case, in order to satisfy equation (\ref{NV1}) we take $P_0^{12}(z)=P_0^{21}(z)=0$ and $P_0^{22}(z)=z^2P_0^{11}(z)$.
Then (\ref{AB}) yields
\begin{eqnarray}
A(\frac{1}{z})=A(z),\quad B(\frac{1}{z})=z^2B(z).
\label{LCDNLS1}
\end{eqnarray}
Analogously, we find $P_N^{12}(z)=P_N^{21}(z)=0$ and $P_N^{11}(z)=z^2P_N^{22}(z)$, which in turn yields
\begin{eqnarray}
A_N(\frac{1}{z})=A_N(z),\quad B_N(\frac{1}{z})=\frac{1}{z^2}B_N(z).
\label{LCDNLS2}
\end{eqnarray}
Letting $z\mapsto \frac{1}{z}$ in the global relation (\ref{grsim2}) and using the symmetries (\ref{LCDNLS1}) and (\ref{LCDNLS2}), we find
\begin{eqnarray}
\begin{split}
z^2a(\frac{1}{z})\rho(z)+\nu z^{2N}b^*(z^*)\rho(z)\rho_1(z)-z^{2(N-1)}a^*(z^*)\rho_1(z)-b(\frac{1}{z})=0,
\quad z\in \bar{D}_{+}.
\end{split}
\label{grsim3}
\end{eqnarray}
Equation (\ref{grsim2}) together with (\ref{grsim3}) yield the expressions for the ratios $\rho(z)$ and $\rho_1(z)$.
Indeed, using (\ref{grsim2}) to eliminate the term $\rho(z)\rho_1(z)$ in (\ref{grsim3}), we obtain
\begin{eqnarray}
\begin{split}
f(z)\rho_1(z)=g(z)\rho(z)-h(z),
\end{split}
\label{ABABNr}
\end{eqnarray}
where
\begin{eqnarray}
\begin{split}
&f(z)= a^*(\frac{1}{z^*})b^*(z^*)-z^{-2}a^*(z^*)b^*(\frac{1}{z^*}),
\\
&g(z)=z^{2N} a(z)b^*(z^*)-z^{-2(N-1)}a(\frac{1}{z})b^*(\frac{1}{z^*}),
\\
&h(z)=z^{2N} b(z)b^*(z^*)-z^{-2N}b(\frac{1}{z})b^*(\frac{1}{z^*}).
\end{split}
\label{fgh}
\end{eqnarray}
Substituting (\ref{ABABNr}) into (\ref{grsim2}), we find
\begin{eqnarray}
\nu b^*(\frac{1}{z^*})g(z)\rho^2(z)+\big(z^{2N}a(z)f(z)-\nu b^*(\frac{1}{z^*})h(z)-a^*(\frac{1}{z^*})g(z)\big)\rho(z)+a^*(\frac{1}{z^*})h(z)-z^{2N}b(z)f(z)=0.
\label{ABs}
\end{eqnarray}
This equation yields the following expression for the ratio $\rho(z)$:
\begin{eqnarray}
\begin{split}
\rho(z)=\frac{1}{2\nu b^*(\frac{1}{z^*})g(z)} \left(a^*(\frac{1}{z^*})g(z)+\nu b^*(\frac{1}{z^*})h(z)-z^{2N}a(z)f(z)
\pm \sqrt{\Delta}\right),
\end{split}
\label{rhos}
\end{eqnarray}
where
\begin{eqnarray*}
\Delta=\big(z^{2N}a(z)f(z)-\nu b^*(\frac{1}{z^*})h(z)-a^*(\frac{1}{z^*})g(z)\big)^2-4\nu b^*(\frac{1}{z^*})g(z)\big(a^*(\frac{1}{z^*})h(z)-z^{2N}b(z)f(z)\big).
\label{Delta}
\end{eqnarray*}
Moreover, equation (\ref{ABABNr}) yields
\begin{eqnarray}
\begin{split}
\rho_1(z)=\frac{g(z)}{f(z)}\rho(z)-\frac{h(z)}{f(z)}.
\end{split}
\label{rho1s}
\end{eqnarray}
Hence, we have expressed the ratios $\rho(z)$ and $\rho_1(z)$ and thus the associated RH problem only in terms of known spectral functions $a(z)$ and $b(z)$.

We note that in the case of the problem on the ``half-line'',
by using algebraic manipulations of the global relation, the ratio $B(z)/A(z)$ can be computed via a simple linear algebraic equation (see \cite{F3,F4,XiaFokas});
while in the case of the problem on the ``interval'', the analogous equations determining the ratios $B(z)/A(z)$ and $B_N(z)/A_N(z)$ involve quadratic nonlinearity (see equations  (\ref{grsim2}), (\ref{grsim3}) and the resulting equation (\ref{ABs})). 
As a result, the expression of the ratios $B(z)/A(z)$ and $B_N(z)/A_N(z)$ in terms of initial data for the problem on the ``interval''
is much complicated than the corresponding expression for the problem on the ``half-line''.

Here we have identified the linear boundary conditions and have shown how to express the RH problem in terms of known spectral functions with such boundary conditions.
It is interesting to study the construction of soliton solutions by using these linearizable boundary conditions.
We will study this topic elsewhere.

We note that different aspects of linearizable boundary
conditions for integrable nonlinear differential-difference equations have been studied in \cite{S,IB1,IB2}.

\section {Conclusions and discussions}\label{section6}
We have shown how to implement the Fokas method to analyse initial-boundary value problems
for DNLS equation on a finite set of integers.
We first presented a Lax pair for the DNLS equation which is convenient for performing the simultaneous spectral analysis.
Then by performing the spectral analysis to the Lax pair, we expressed the solution of the DNLS equation in terms of the solution of an associated matrix
Riemann-Hilbert problem in the complex $z$-plane of the spectral parameter. The jump matrix of this Riemann-Hilbert problem has explicit $(n,t)$-dependence
and it involves six spectral functions:  $\left\{a(z),b(z)\right\}$, $\left\{A(z),B(z)\right\}$ and $\left\{A_N(z),B_N(z)\right\}$.
The spectral functions $\left\{a(z),b(z)\right\}$ depend on the initial value: $q_{0}(n)$;
whereas the spectral functions $\left\{A(z),B(z)\right\}$ and $\left\{A_N(z),B_N(z)\right\}$ depend on two sets of boundary values: $\{g_{-1}(t),g_0(t)\}$ and $\{g_{N-1}(t),g_N(t)\}$.
We showed that the spectral functions satisfy an algebraic global
relation and characterized the unknown boundary values in terms of the given initial and
boundary values by employing this global relation.
We also discussed the linearizable boundary conditions.

The approach presented here can be generalized to other integrable discrete nonlinear evolution equations on finite sets of integers,
such as the Volterra lattice \cite{V1,V2}, the Toda lattice \cite{T1} and the four-potential Ablowitz-Ladik lattice \cite{AL1,Geng2,ZC,ZC2}.

\section*{ACKNOWLEDGMENTS}
The author thanks the referees for their valuable comments.
The author also thanks A.S. Fokas for helpful suggestions and thanks the Department of Applied Mathematics and Theoretical Physics, University of Cambridge, for the kind hospitality.
This work was supported by the National Natural Science Foundation of China (Grant No. 11771186)
and by the Jiangsu Government Scholarship for Overseas Studies.

\vspace{1cm}
\small{

}
\end{document}